\RequirePackage{fix-cm}
\documentclass[smallextended,numbook]{svjour3}

\smartqed  
\usepackage{graphicx}
\usepackage{pifont}

\newcommand{\ignore}[1]{}

\newcommand{\reviseLiu}[1][\textcolor{black}]{#1}

\newcommand{\TODO}[1][\textcolor{black}]{#1}

\usepackage{colortbl}

\usepackage[colorlinks=true,linkcolor=blue,citecolor=blue,urlcolor=blue]{hyperref}

\usepackage{url}
\usepackage{breakurl}
\usepackage{natbib}

\usepackage{booktabs}

\usepackage{tabulary}

\usepackage{cleveref}
\usepackage{threeparttable}

\usepackage{multirow}

\usepackage{tabularx}
\usepackage{adjustbox}

\usepackage{makecell}
\usepackage{chngcntr}
\counterwithout{table}{section}
\counterwithout{figure}{section}   
\usepackage{subfigure}

\usepackage{listings}
\usepackage{framed}
\usepackage[framemethod=TikZ]{mdframed}

\usepackage{wasysym}

\usepackage{enumitem}
\usepackage{amssymb} 

\definecolor{KFTitle}{HTML}{3A3A3A} 
\definecolor{KFBack}{HTML}{F2F2F2}  

\usepackage[skins,breakable]{tcolorbox}
\usepackage{enumitem} 
\newtcolorbox{keyfindingbox}[1]{%
  breakable,
  enhanced,
  colback=KFBack,
  boxrule=1.0pt,
  arc=5pt,
  title={#1},
  colbacktitle=KFTitle,
  coltitle=white,
  fonttitle=\bfseries,
  boxed title style={%
    boxrule=0pt,
    arc=5pt,
    left=8pt,right=8pt,
    top=4pt,bottom=4pt
  },
  left=8pt,right=8pt,
  top=8pt,bottom=8pt,
  overlay unbroken={\draw[rounded corners=5pt,line width=1.0pt]
    (frame.south west) rectangle (frame.north east);},
  overlay first={\draw[rounded corners=5pt,line width=1.0pt]
    (frame.south west) rectangle (frame.north east);},
  overlay middle={\draw[line width=1.0pt,color=KFFrame]
    (frame.south west) rectangle (frame.north east);},
  overlay last={\draw[rounded corners=5pt,line width=1.0pt]
    (frame.south west) rectangle (frame.north east);},
    title after break={},
}

\usepackage[switch]{lineno}

\usepackage{marvosym}

\usepackage{microtype}
\begin{document}

\title{Rethinking Software Misconfigurations in the Real World: An Empirical Study and Literature Analysis
}

\titlerunning{Rethinking Software Misconfigurations in the Real World}        

\author{Yuhao Liu\textsuperscript{1} \and
        Yingnan Zhou\textsuperscript{1} \and
        Hanfeng Zhang\textsuperscript{1} \and
        Zhiwei Chang\textsuperscript{1} \and
        Sihan Xu\textsuperscript{1} \and
        Yan Jia\textsuperscript{1} \and
        Wei Wang\textsuperscript{2} \and
        Juncheng Hu\textsuperscript{3} \and
        Zheli Liu\textsuperscript{1}
}

\authorrunning{Yuhao Liu et al.} 

\institute{ Yuhao Liu \at
              \email{yuhao.liu@mail.nankai.edu.cn}
           \and
           Yingnan Zhou \at
              \email{yingnan.zhou@mail.nankai.edu.cn}
           \and
           Hanfeng Zhang \at
              \email{zhanghf2048@gmail.com} 
           \and
           Zhiwei Chang \at
              \email{18851093355@163.com}
           \and
           \Letter~Sihan Xu \at
              \email{xusihan@nankai.edu.cn}
           \and
           \Letter~Yan Jia \at
              \email{jiay@nankai.edu.cn}
           \and
           Wei Wang \at
              \email{wei.wang@xjtu.edu.cn}
           \and
           Juncheng Hu
           \at
              \email{jchu@jlu.edu.cn}
            \and
           Zheli Liu \at
              \email{liuzheli@nankai.edu.cn}\\\\
\textsuperscript{1}~~~Key Laboratory of Data and Intelligent System Security, Ministry of Education, China (DISSec), Tianjin Key Laboratory of Network and Data Security Technology (NDST), College of Cryptology and Cyber Science, Nankai University, Tianjin 300350, China \\
\textsuperscript{2}~~~MOE KLINNS Lab, Xi’an Jiaotong University, Shaanxi, China \\
\textsuperscript{3}~~~College of Computer Science and Technology, Jilin University, Jilin, China
}

\date{Received: date / Accepted: date}

\maketitle


\begin{abstract}
Software misconfiguration has consistently been a major reason for software failures. Over the past two decades, much work has been done to detect and diagnose software misconfigurations. However, there is still a gap between real-world misconfigurations and the literature. It is desirable to investigate whether existing taxonomy and tools are applicable for real-world misconfigurations in modern software. In this paper, we conduct an empirical study on 772 real-world misconfiguration issues, based on which we propose a novel classification of the root causes of software misconfigurations, i.e., constraint violation, resource unavailability, component integration error, and configuration semantic misinterpretation. 
Then, we systematically review the literature on misconfiguration troubleshooting to study the trends of research and the practicality of the tools and datasets in this field. We find that the research targets have changed from system and infrastructure software to advanced applications (e.g., cloud service). Meanwhile, research on non-crash misconfigurations has also grown significantly. 
Despite the progress, a majority of studies lack reproducibility due to the unavailable tools and evaluation datasets. In total, only eleven tools and four datasets are publicly available. We analyze the trends of existing literature on misconfiguration troubleshooting, summarize the challenges that users are faced with, and highlight the suggestions to mitigate and diagnose software misconfigurations. We release the real-world dataset of misconfiguration issues for follow-up research. 
\keywords{Software misconfigurations, Empirical study, Literature analysis, Configuration management, Troubleshooting}
\end{abstract}

\section{Introduction}\label{sec1}

Modern software systems are equipped with an increasing number of configuration options. Using different configuration settings, software systems allow customization of functional behaviors that are tailored to support different user intentions.
Although such a mechanism provides the flexibility of software systems, it introduces configuration errors (i.e., misconfigurations), which may lead to severe consequences. For example, in February 2023, a misconfiguration in a Microsoft cloud email server caused data of 20,600 accounts to be exposed without password protection\footnote{ServiceNow misconfiguration went unexploited, but still cause for concern, \url{https://www.scmagazine.com/news/servicenow-misconfiguration-went-unexploited-but-still-cause-for-concern}}. 

Software misconfiguration has consistently been a major factor in software failures in large-scale systems\footnote{Hacking Guatemala’s DNS -- Spying on Active Directory Users By Exploiting a TLD Misconfiguration, \url{https://thehackerblog.com/hacking-guatemalas-dns-spying-on-active-directory-users-by-exploiting-a-tld-misconfiguration/index.html}}\textsuperscript{,}\footnote{2020 Data Breach Investigations Report, \url{https://www.verizon.com/business/resources/reports/2020-data-breach-investigations-report.pdf}}~(Xu et al.~\citeyear{Empirical2015Xu}). 
Some studies aim to methodically understand real-world software misconfigurations by categorizing them according to their fundamental causes or consequences.
From the perspective of consequences, Zhang et al.~(\citeyear{2013confDiagnoser,2014confSuggester}) distinguished software misconfigurations by whether they can lead to system crashes. Yin et al.~(\citeyear{yin2011empirical}) presented a detailed classification of misconfigurations: {parameter}, {compatibility}, and {component} issues, where \textit{parameter} refers to parameter mistakes, {including ``\textit{incorrect format}'' and ``\textit{unsupported value}'',} \textit{compatibility} refers to the compatibility among different components or modules, and \textit{component} refers to ``\textit{neither parameter mistakes nor compatibility problems}''. However, these classifications provided limited assistance in understanding fundamental causes. For example, a hardware limitation issue (\textit{RC$\mathit{_{II}}$-Case\#210}\footnote{\textit{RC$\mathit{_{II}}$} indicates the second root cause category in~\Cref{root_cause} and its case ID in our dataset is 210.}) occurred when the system loaded a large dataset and ran out of memory because the configuration option \texttt{memory\_limit} was configured to a very low value. However, this case would be incorrectly classified as component misconfigurations according to the classification proposed by Yin et al.~(\citeyear{yin2011empirical}). Since the configured value is legal, it is not classified as a parameter mistake according to the definition. 

In addition to classification, many approaches have been proposed to troubleshoot software misconfigurations, such as statistical analysis~(Wang et al.~\citeyear{2003peerpressure}; Attariyan and Flinn~\citeyear{2008sigConf}; Zhang and Ernst~\citeyear{2013confDiagnoser}), program static analysis~(Zhang and Ernst~\citeyear{2013confDiagnoser}, \citeyear{2014confSuggester}; Dong et al.~\citeyear{2015confDoctor}; Zhang et al.~\citeyear{zhang2021configX}), instrumentation~(Su et al.~\citeyear{2007autobash}; Attariyan et al.~\citeyear{2010confAid}, \citeyear{2012Xray}), machine learning~(Santolucito et al.~\citeyear{2016configc}, \citeyear{2017configv}; Xiang et al.~\citeyear{2020pracextractor}), etc. With the emergence of many modern software, it is necessary to investigate whether existing methods for troubleshooting configuration errors are still effective in resolving real-world configuration issues. To better understand configuration pitfalls and mitigate misconfiguration issues, we conduct an empirical study on misconfigurations in the real world and analyze the existing literature on software misconfiguration troubleshooting. This paper focuses on software misconfigurations that arise in real-world deployment and runtime settings, typically when users or developers configure software systems. We do not consider internal implementation defects, including configuration implementation bugs and software bugs, which may be triggered when the supplied configuration values are valid. Specifically, our research focuses on four research questions. \textbf{RQ1}: What are the root causes of {real-world} software misconfigurations? \textbf{RQ2}: What are the progress and trends of existing literature {on misconfiguration troubleshooting}? \textbf{RQ3}: How practical are the existing tools and datasets in addressing software misconfigurations? \textbf{RQ4}: What challenges have yet to be solved in detecting and diagnosing real-world misconfigurations? 

To address \textbf{RQ1}, we first summarize the life-cycle of software configuration. Then, we collect 772 real-world misconfiguration issues from two online communities and three official customer service channels, based on which we present a novel and systematic classification of the root causes of misconfigurations. It took two man-months to manually {select} and analyze 168,054 user-raised issues from these sources. Based on the dataset, we summarize the main root causes by four types, i.e., \textit{constraint violation (RC$\mathit{_{I}}$)}, \textit{resource unavailability (RC$\mathit{_{II}}$)}, \textit{component integration error (RC$\mathit{_{III}}$)}, and \textit{configuration semantic misinterpretation (RC$\mathit{_{IV}}$)}, and further divide each root cause into several subtypes. To address \textbf{RQ2}, we conduct a comprehensive review of the literature on misconfiguration troubleshooting from 2003 to 2024. We investigate the field of software misconfiguration research in five aspects: research targets, misconfiguration symptoms, root causes, troubleshooting techniques, and debugging artifacts. We find that the attention of researchers evolves from incorrect {functionalities} caused by misconfigurations to implicit silent issues (e.g., performance {degradation}). The evaluation targets also move towards large-scale complicated distributed systems. To address \textbf{RQ3}, we first investigate existing tools for misconfiguration troubleshooting from the aspects of their availability, maintainability, and adaptability. Then, we discuss the benchmarks that can be used to evaluate the effectiveness of misconfiguration troubleshooting tools. We find that only \TODO{11} tools and 4 datasets in this field are publicly available. Furthermore, among the 11 existing tools, 7 tools lack maintenance and 9 tools are tightly coupled with their research targets. To address \textbf{RQ4}, we discuss the important challenges that hinder software misconfiguration troubleshooting. Our analysis reveals that insufficient feedback messages affect 14.51\% (112/772) of real-world cases. Existing techniques still heavily rely on manual efforts for modeling and labeling.

In summary, we make the main contributions as follows.
\begin{itemize}

    \item \textbf{Taxonomy.} We manually collected and analyzed 772 real-world misconfigurations raised by users, based on which we present a novel and systematic classification of the root causes of software misconfigurations.
    
    \item \textbf{Dataset.}  We make the dataset publicly available:
    {\url{https://github.com/anabioticsoul/misconfiguration_datasets}}. To the best of our knowledge, the dataset is the most comprehensive dataset about real-world misconfigurations. The dataset can serve as a real-world benchmark to aid the research on software misconfiguration.

    \item \textbf{Insights.} We present the first in-depth analysis on software misconfiguration from the perspectives of both real-world cases and the literature. We investigate the research trends in software misconfiguration, discuss the practicality of existing tools and benchmarks in the field, and summarize the challenges and suggestions to facilitate the follow-up research. 
    
\end{itemize}

\section{Life-cycle of Software Configuration}\label{sec2}

\begin{figure*}[ht]
\centering 
\includegraphics[width=0.95\textwidth]{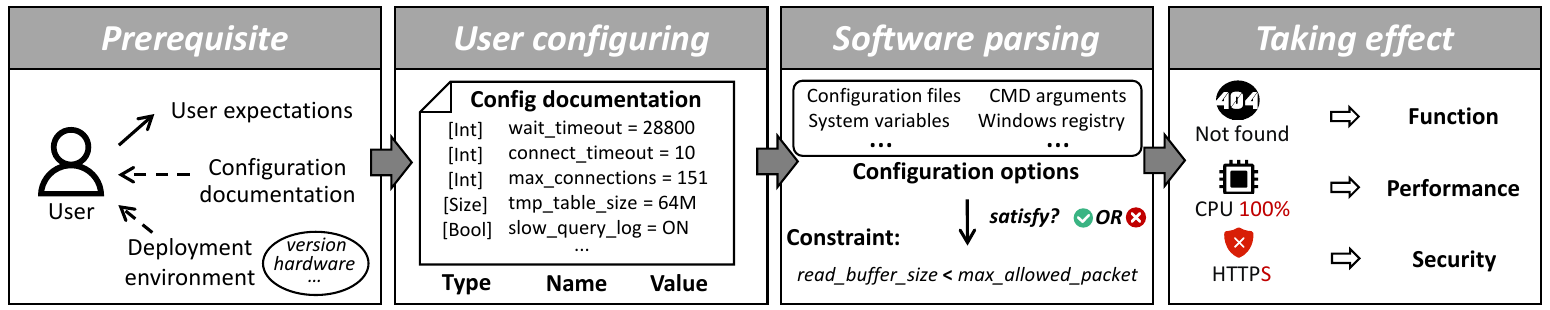} 
\caption{Overview of software configuration life-cycle} 
\label{ConfOverview} 
\end{figure*}

Configuration is an important mechanism for customizing functional features. It allows users to modify the functionalities of a configurable software system. 
To thoroughly analyze the fundamental reasons behind software misconfigurations, we outline the life-cycle of software configuration and explore the potential reasons that may cause configuration errors {in each stage}. Generally, as depicted in Figure~\ref{ConfOverview}, the life-cycle of software configuration includes four stages, i.e., prerequisite, user configuring, software parsing and taking effect.

\noindent\textbf{Prerequisite.} To make software more flexible, developers usually make some functional features customizable by allowing users to adjust the relevant configuration options. By learning the relevant documentation or demonstrations provided by developers, users (e.g., customers and system operators) can modify the software settings according to their deployment environment and requirements. In summary, configuration documentation, user expectations, and the deployment environment are three key factors that must be considered in the preparation phase.

\noindent\textbf{User configuring.} To customize functional features, users can adjust configuration options before starting the software. Typically, a configuration option is designed to be a key-value pair. Configuration option names are typically readable strings that encapsulate the option's purpose. For example, the configuration \texttt{query\_cache\_size} in MySQL determines the amount of memory allocated for caching query results. Accordingly, the value of the option specifies the user's choice and affects the operation of the software. The common types of option values include integer, boolean, string, etc. Typically, the settings for configuration options are bound by specific rules determined by the meaning of the options. For instance, in Apache Hadoop, the option \texttt{hadoop.tmp.dir} specifies a certain directory and its value should be a URI. 
However, options of different software may follow different rules, making the user configuration process error-prone, e.g., the parsing strategies in MySQL and PHP {(\textit{RC$\mathit{_{I}}$-Case\#35} and \textit{RC$\mathit{_{I}}$-Case\#2} in our dataset).}

\noindent\textbf{Software parsing.} Software reads and parses configurations from various sources such as configuration files, system variables, command-line arguments and the windows registry. 
Configurations can be read at any time. Some applications like PyHive and Netty read and check configurations at initialization, while others like Apache Hadoop may configure options before or after the software starts. 
After reading the values of configuration options, the software checks if they meet specific constraints. 
If {the values of some }configurations are missing or do not satisfy the parsing requirements, the software may produce error messages. 
Sometimes, the software may use default values to replace missing or invalid settings. 

\noindent\textbf{Taking effect.} Configurations determine the functionality and performance of a software product.
{A configuration may take effect from three aspects, i.e., functionality, performance, and security. 
Users need to properly set configuration options to make software behave as expected. For example, users may customize software functionalities tailored to their own business requirements or configure software to satisfy the performance requirements they need. In addition, {software configuration may also affect the security of software, e.g., configurations related to authentication and access control policy}.} 
{However, due to the complexity of configurations, users may unintentionally set wrong values to configuration options, leading to unexpected software behaviors such as crashes, low performance, and security risks}.

\section{Methodology}\label{sec3}

\subsection{Overview}\label{sec3:1}
\begin{figure*}[htbp]
\centering
\includegraphics[width=0.9\textwidth]{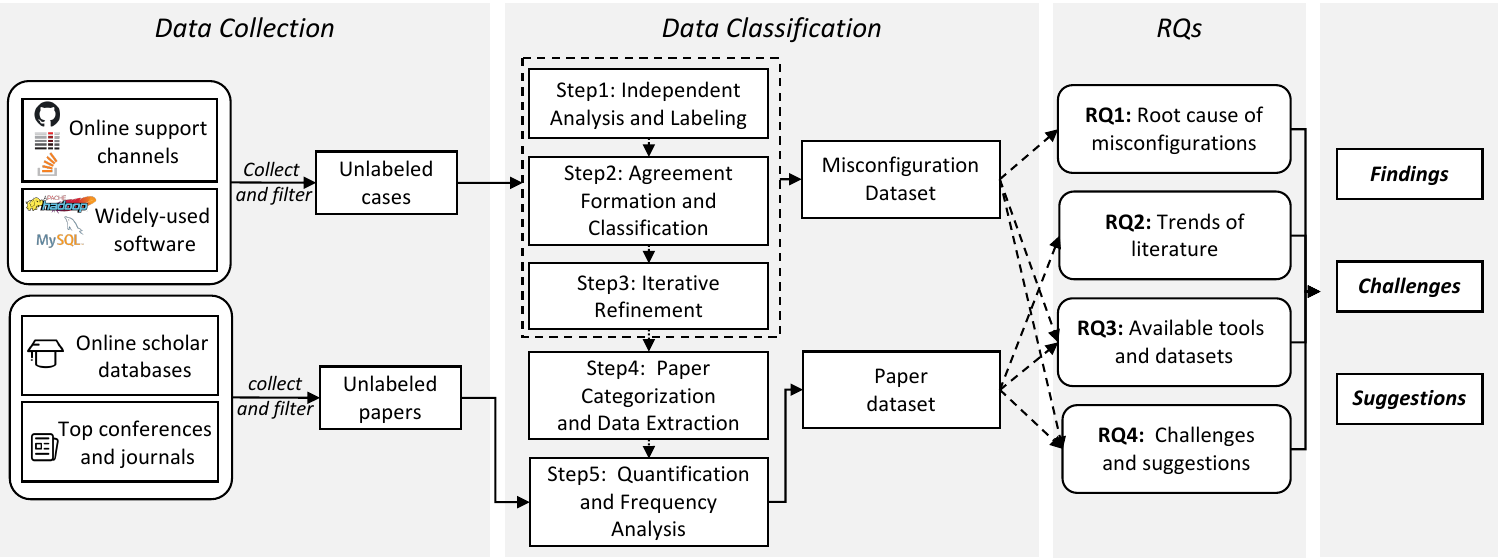}
\caption{Overview of our methodology}
\label{work_flow}
\end{figure*}

To investigate the root causes of real-world software misconfigurations and trends in the literature on misconfiguration troubleshooting, we present a two-phase approach as shown in Figure~\ref{work_flow}. In the data collection phase, we first searched with configuration-related keywords and collected 168,054 issues from online communities such as StackOverflow. Then, we filtered out invalid cases and obtained 772 real-world misconfiguration cases~(\Cref{sec3:2}). Moreover, to review the research on software misconfiguration, we also collected 60 academic papers on misconfiguration detection and diagnosis~(\Cref{sec3:3}). In the data classification phase, we carefully inspected each case and summarized four types of root causes of misconfigurations. Then, we classified the papers according to the proposed taxonomy and extracted four aspects to quantify the evolutionary trends of the literature~(\Cref{sec3:4}). The results we collected are provided at the end of each subsection.

\vspace{-1.5\baselineskip}

\subsection{Real-world misconfiguration collection}\label{sec3:2}
To methodically understand real-world software misconfigurations, our targets include popular software (e.g., MySQL\footnote{\url{https://www.mysql.com/}}, PHP\footnote{\url{https://www.php.net/}}, Apache httpd\footnote{\url{https://httpd.apache.org/}}, Nginx\footnote{\url{https://nginx.org/}}, PostgreSQL\footnote{\url{https://www.postgresql.org/}}, and Hadoop\footnote{\url{https://hadoop.apache.org/}}) that have been widely used by previous studies~(Attariyan and Flinn~\citeyear{2010confAid}; Yin et al.~\citeyear{yin2011empirical}; Attariyan et al.~\citeyear{2012Xray}; Zhang et al.~\citeyear{2014Encore}; Xiang et al.~\citeyear{2020pracextractor}). Each software is highly configurable, with its source code, configuration documentation, and community discussions publicly accessible, which makes it easy to identify the root causes of misconfigurations. We collected the misconfiguration cases from two online communities (i.e., StackOverflow\footnote{\url{https://stackoverflow.com/}} and ServerFault\footnote{\url{https://serverfault.com/}}) and {three} official customer service channels (i.e., GitHub\footnote{\url{https://github.com/}}, Mailing lists\footnote{\url{https://lists.apache.org/}}, and Official online forums). These active technical communities online have millions of users, covering a wide range of real-world issues, and are widely used in studies in misconfiguration troubleshooting~(Barua et al.~\citeyear{2014stackoverflow}; Yin et al.~\citeyear{yin2011empirical}; Wang et al.~\citeyear{wang2024exploratory}).

To obtain real-world misconfiguration cases, we defined the selection criteria by incorporating the following steps:
\begin{itemize}
    \item \textbf{Step~1:} We selected the target software name as a search tag and used configuration-related keywords, i.e., {\textit{configuration*}, \textit{misconfiguration*}, \textit{configure*}, \textit{configuration error*}, and \textit{configuration fault*}.} To retrieve relevant issues reported by users, we constructed boolean search expressions combining the software tag with these keywords. 

    \item \textbf{Step~2:} We only selected issues whose answers had already been accepted by the questioners and whose problems were caused by misconfigurations that could be resolved through configuration changes. Such issues provide clearer evidence that the failures were configuration-related rather than implementation defects, helping identify the root causes of configuration errors.

\end{itemize}

\begin{table}[t]
\footnotesize
\centering
\caption{Filtering quantity of collected issues by criteria}
\label{sample-used}
\begin{tabular}{cccc}
\toprule
\textbf{Software} & \textbf{Total}  & \textbf{Step~1} & \textbf{Step~2}\\
\midrule
MySQL         & 25435    & 391    & 115 \\
PHP           & 41468    & 164    & 79  \\
Apache httpd  & 44146    & 510    & 171 \\
Nginx         & 24433    & 405    & 203 \\
PostgreSQL    & 7287     & 117    & 54  \\
Hadoop        & 7566     & 100    & 42  \\
Others        & 17719    & 626    & 108 \\
Total         & 168054   & 2313   & 772 \\
\bottomrule
\end{tabular}
\end{table}

\noindent\textbf{Results.}
Five authors participated in the collection and verification of misconfiguration issues. As shown in \Cref{sample-used}, we collected a total of 168,054 issues from seven popular software systems. Using the predefined keywords, we identified 2,313 misconfiguration-related issues. We then manually checked these issues and filtered out those that remained unresolved or could not be fixed by modifying configuration settings. All five authors were involved in the manual screening of the 2,313 candidate issues. Each issue was checked by two evaluators independently, and disagreements were resolved through discussion. To evaluate the reliability of this screening process, we calculated a Cohen's $\kappa$ of 0.94 for this process, indicating substantial agreement among the evaluators. By this means, we obtained a dataset of 772 real-world misconfiguration cases.

\subsection{Literature collection}\label{sec3:3}
Following the guidelines of systematic literature reviews by Kitchenham~et al.~(\citeyear{2007SLRguidelines}), we adopted a search strategy to ensure comprehensive coverage of relevant studies. 

\noindent\textbf{Search sources.} To collect relevant academic publications, we selected several digital libraries as search sources, including CiteSeer, IEEE Xplore, ACM Digital Library, ScienceDirect, and Wiley Online Library. These repositories cover a broad spectrum of high-quality peer-reviewed research in computer science and software engineering, ensuring the completeness and credibility of our literature survey.

\noindent\textbf{Search strategy.} We initially performed a search on 29 top-tier conferences and journals in the domains of software engineering and information security, as listed in \Cref{publications}. These venues were selected based on their high impact and wide recognition in the research community, which are known to publish state-of-the-art studies related to software misconfiguration troubleshooting. We crawled the papers from the top venues from December 2003 to September 2024. The keyword set includes \textit{configuration*}, \textit{misconfiguration*}, \textit{configure*}, \textit{configuration error*}, and \textit{configuration fault*}. To complement our keyword-based search, we performed both backward and forward ``snowballing" on the initially selected papers~(Wohlin~\citeyear{2014snowballing}), examining their reference lists and citations to identify additional candidates. This cycle continued until no further relevant papers satisfying our inclusion criteria were discovered.

\noindent\textbf{Selection criteria.} In order to determine the relevance of the collected papers, we followed a structured manual review process. \reviseLiu{Specifically, a paper was considered relevant if it presented techniques or tools focused on software misconfiguration troubleshooting in deployment and runtime settings, e.g., the detection, diagnosis, or resolution of configuration errors. Note that we excluded papers focusing only on configuration implementation bugs and software bugs~(e.g., configuration bug detection or  misconfiguration injection). We also excluded papers focusing only on configuration dependency analysis, benchmark generation, or non-software domains.} During the screening process, we first reviewed the title and abstract of each paper. If the relevance remained unclear, we examined the introduction and methodology sections. To ensure the reliability of our literature collection and minimize subjective bias, we adopted a double-review process. Specifically, two authors independently screened each paper according to inclusion and exclusion criteria. After independent review, we compared the results and measured inter-rater agreement with Cohen’s $\kappa$. Any disagreements were resolved through discussion, and when necessary, the third author acted as an arbitrator.

\begin{table*}[t]
\footnotesize
\centering
\caption{Publications venues included in search strategy}
\label{publications}
\resizebox{\hsize}{!}{
\begin{tabular}{rl}
\toprule
\textbf{Acronym} & \textbf{Venues} \\
\midrule
ASE & International Conference on Automated Software Engineering \\
ASPLOS & International Conference on Architectural Support  for Programming Languages and Operating Systems \\
CCS & ACM Conference on Computer and Communications Security \\
ESE & Empirical Software Engineering \\
ESEC/FSE & ACM Joint European Software Engineering Conference and Symposium on the Foundations of Software Engineering \\
EuroSys & European Conference on Computer Systems \\
ICPC & IEEE International Conference on Program Comprehension \\
ICSE & International Conference on Software Engineering \\
IETS & IET Software \\
IST & Information and Software Technology \\
ISSTA & International Symposium on Software Testing and Analysis \\
JFP & Journal of Functional Programming \\
JSS & Journal of Systems and Software \\
NSDI & Symposium on Network System Design and Implementation \\
OOPSLA & Conference on Object-Oriented Programming Systems, Languages, and Applications \\
OSDI & USENIX Symposium on Operating Systems Design and Implementation \\
RE & Requirements Engineering \\
SCP & Science of Computer Programming \\
SOSP & ACM Symposium on Operating Systems Principles \\
SoSyM & Software and Systems Modeling \\
SPE & Software: Practice and Experience \\
SPLC & International Systems and Software Product Line Conference \\
STVR & Software Testing, Verification and Reliability \\
TDSC & IEEE Transactions on Dependable and Secure Computing \\
TOPLAS & ACM Transactions on Programming Languages and Systems \\
TOSEM & ACM Transactions on Software Engineering and Methodology \\
TSC & IEEE Transactions on Services Computing \\
TSE & IEEE Transactions on Software Engineering \\
USENIX ATC & USENIX Annual Technical Conference \\
\bottomrule
\end{tabular}
}
\end{table*}

\noindent\textbf{Results.}
Five authors independently conducted the automated search and manual verification. First, each author selected three publication venues to crawl papers from the digital libraries, ensuring coverage of \Cref{publications}. In total, we systematically crawled 47,303 papers. Using a predefined set of keywords, we filtered these papers and obtained 367 candidate papers. Second, we applied forward and backward ``snowballing'' on citation relationships and identified 223 additional papers. After merging the results and removing duplicates, 104 papers remained. \reviseLiu{Third, as a complement to the automated search, we incorporated expert recommendations. These candidate papers were evaluated against our selection criteria, resulting in the inclusion of two relevant papers.} Finally, all 473 candidate papers were manually screened to determine whether they addressed misconfiguration troubleshooting. To assess the reliability of the screening process, we calculated Cohen's $\kappa$ for the inclusion decisions, which was 0.72, indicating substantial agreement among the evaluators. After resolving disagreements through discussion, 60 papers were finally included in our study. The replication package is publicly available\footnote{\url{https://github.com/anabioticsoul/misconfiguration_datasets}}.

\subsection{Systematic analysis}\label{sec3:4}
To identify the root causes of software misconfigurations and summarize the trends of the existing literature, we employed the widely recognized methodologies of open coding~(Creswell and Poth~\citeyear{creswell2016}; Wang et al.~\citeyear{wang2021exploratory}) and systematic mapping-style analysis~(Petersen et al.~\citeyear{2015SMSguidelines}) to analyze our collected datasets. Five computer science graduates with bachelor's degrees participated in a manual analysis process, requiring a total effort of two person-months. As shown in~\Cref{work_flow}, we followed a five-step evaluation approach to conduct the study.

\noindent\textbf{Step 1: Independent analysis and label}.
In the first step, the collected cases were distributed according to the target software systems. To align the evaluation criteria, each evaluator analyzed one or two target software systems, ensuring that every software system was initially assessed by a primary evaluator. They carefully checked all the related misconfigurations. Specifically, they first read the information provided by the questioners and the accepted suggestions in each case. Next, they extracted key information from each case, including trigger conditions, error messages, system feedback, and other contextual information. These extracted attributes were also used to support subsequent analyses, such as understanding the causes of configuration errors and evaluating the quality of system feedback. Based on these details, the evaluators identified the root causes of the misconfigurations and categorized them into specific subtypes. To ensure reliability and mitigate individual bias, every case was independently re-analyzed by a second evaluator who had not participated in the initial analysis of that case. Thus, all cases were double-coded by two evaluators.

\noindent\textbf{Step 2: Agreement formation and classification}.
In the second step, the evaluators compared their labeling results. Cohen's $\kappa$ between two evaluators was {0.88}, indicating substantial agreement based on Landis and Koch's guidelines~(Landis and Koch~\citeyear{1977measurement}). Disagreements were resolved through discussions with a third evaluator until a consensus was reached. The three evaluators then met to refine the classification taxonomy. They clarified the descriptions and boundaries of categories, adjusted the categories, and modified the hierarchical structure.
 
\noindent\textbf{Step 3: Iterative refinement}.
In the third step, after establishing the initial classification scheme, further iterations focused on merging semantically similar labels and clarifying ambiguous definitions. Guided by the established classification and labeling strategies, the evaluators reviewed each case and discussed it. Similar labels are marked as subtypes and combined into one root cause. The discussions continued until all cases were evaluated without disagreement. This step established a robust classification for misconfigurations.

\noindent\textbf{Step 4: Paper categorization and data extraction}.
In the fourth step, evaluators extracted detailed information regarding five aspects for further analysis: misconfiguration symptoms, root causes, research targets, troubleshooting techniques, and debugging artifacts. For the misconfiguration symptoms and root causes, two evaluators first read the benchmarks used by the papers to record the specific configuration errors and bad behaviors. As for papers without explicit statements about the misconfiguration types, the evaluators determined the specific symptoms and root causes addressed by the literature based on their proposed methods and resolution goals. For the remaining three aspects, two evaluators carefully reviewed the ``Methodology'' and ``Evaluation'' sections of each paper and recorded the specific entities being used. After labeling, evaluators discussed any disagreements. The process of discussing the labels of previous works was similar to Step 2.

\noindent\textbf{Step 5: Quantification and frequency analysis}. 
In the final step, we quantified the extracted labels to derive the trends presented in \Cref{sec5}. Since a paper may involve multiple labels within the same aspect, we adopted a multi-label counting strategy at the category level. Specifically, each paper was counted at most once per category within each aspect. For example, for research targets, if a paper evaluated multiple software systems within the same category (e.g., MySQL and PostgreSQL, both databases), it was counted only once in that category. If a paper evaluated software systems belonging to different categories (e.g., MySQL as a database and Hadoop as cloud software), each distinct category was counted once. Note that one paper may contribute to multiple categories.

\begin{figure*}[ht]
\centering
\includegraphics[width=0.90\textwidth]{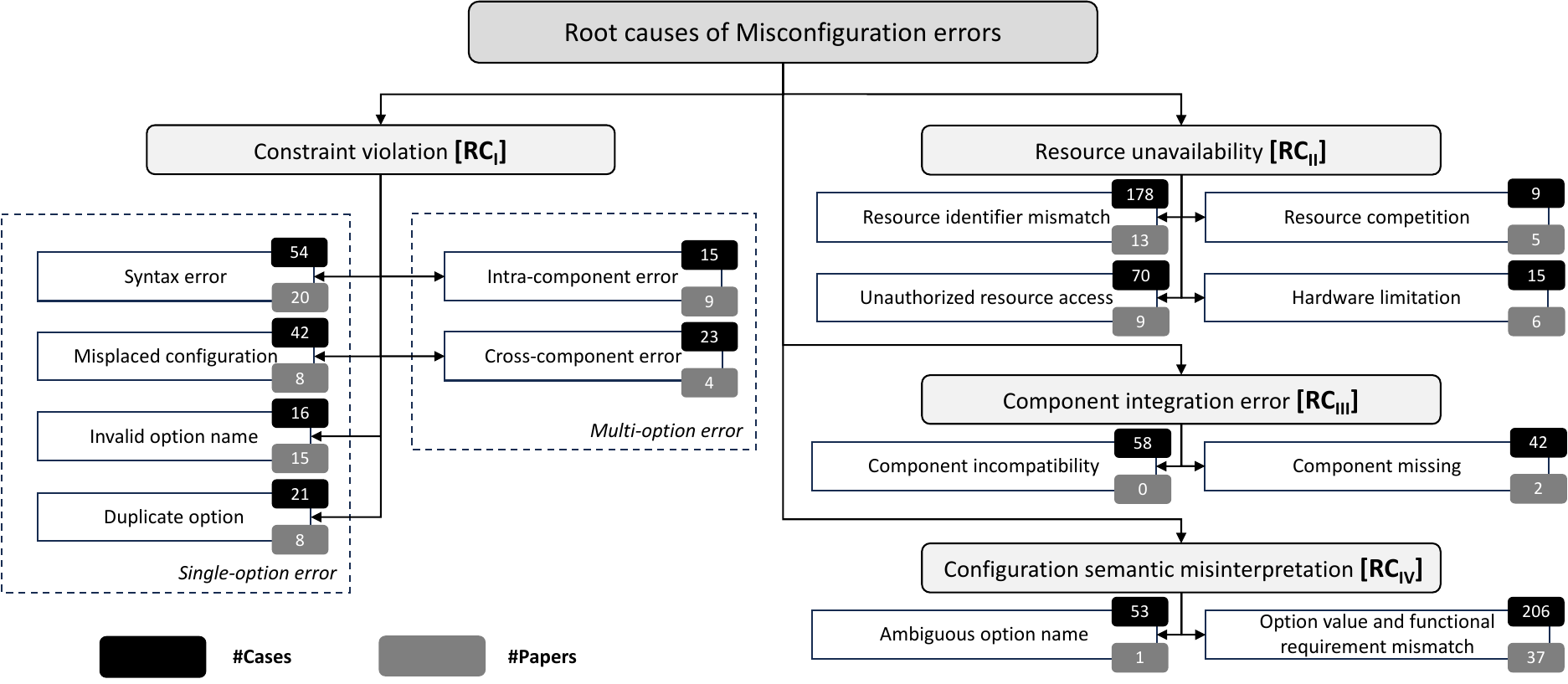}
\caption{Root causes of configuration errors}
\label{root_cause}
\end{figure*}

\noindent\textbf{Results}.
Figure~\ref{root_cause} illustrates the classification of the root causes of software misconfigurations, as well as the numbers of real-world cases and research papers related to each type/subtype of root cause.
In total, we categorized the root causes of misconfigurations into four groups, i.e., {constraint violation}, {resource unavailability}, {component integration error}, and {configuration semantic misinterpretation}. To facilitate the understanding of the root causes, we further divided each group into several subgroups.
In~\Cref{root_cause}, the numbers of misconfiguration cases and related papers are shown in the upper right and bottom right of each type, respectively.
We made the misconfiguration dataset publicly available online\footnote{\url{https://github.com/anabioticsoul/misconfiguration\_datasets}}. In the dataset, each misconfiguration case has its own ID, year, software name, link, description, and the type/subtype of root cause.
In the remainder of this paper, we use \textit{RC$\mathit{_{I}}$}, \textit{RC$\mathit{_{II}}$}, \textit{RC$\mathit{_{III}}$}, and \textit{RC$\mathit{_{IV}}$} with \textit{Case\#\textit{N}} (N is the sequence number in our dataset) to denote the cases whose root causes are {constraint violation}, {resource unavailability}, {component integration error}, and {configuration semantic misinterpretation}, respectively.

\section{RQ1: Root Cause of Misconfigurations}\label{sec4}
{To address RQ1, we adopted widely-used open coding for systematic analysis of the real-world misconfiguration cases. We categorize the root causes of misconfigurations into four groups. This section discusses each root cause in detail, along with real-world cases as examples.}

\subsection{Root Cause 1 (RC$\mathbf{_{I}}$): Constraint violation}\label{rc1}
Constraint violation errors occur when user-specified configuration options do not comply with the constraints predefined by the software developers. The gap between user knowledge and configuration constraints may result in such constraint violation errors. To investigate the reasons behind constraint violation, we further divide it into six subtypes according to the observations of real-world misconfigurations. \reviseLiu{For clarity, we further organize these subtypes into two groups based on whether the violated constraint is caused by only a single configuration option or correlations among multiple options. 
\textit{Single-option error} refers to violations of constraints that are only caused by an individual configuration option, including syntax error, invalid option name, misplaced configuration, and duplicate option. 
\textit{Multi-option error} refers to violations caused by unsatisfied constraints among multiple configuration options, including intra-component error and cross-component error.} Constraint violation accounts for 22.15\% (171/772) of all misconfiguration cases.

\subsubsection{Syntax error}\label{rc1.1}
\begin{figure*}[htp]
\centering
\includegraphics[width=1.0\hsize]{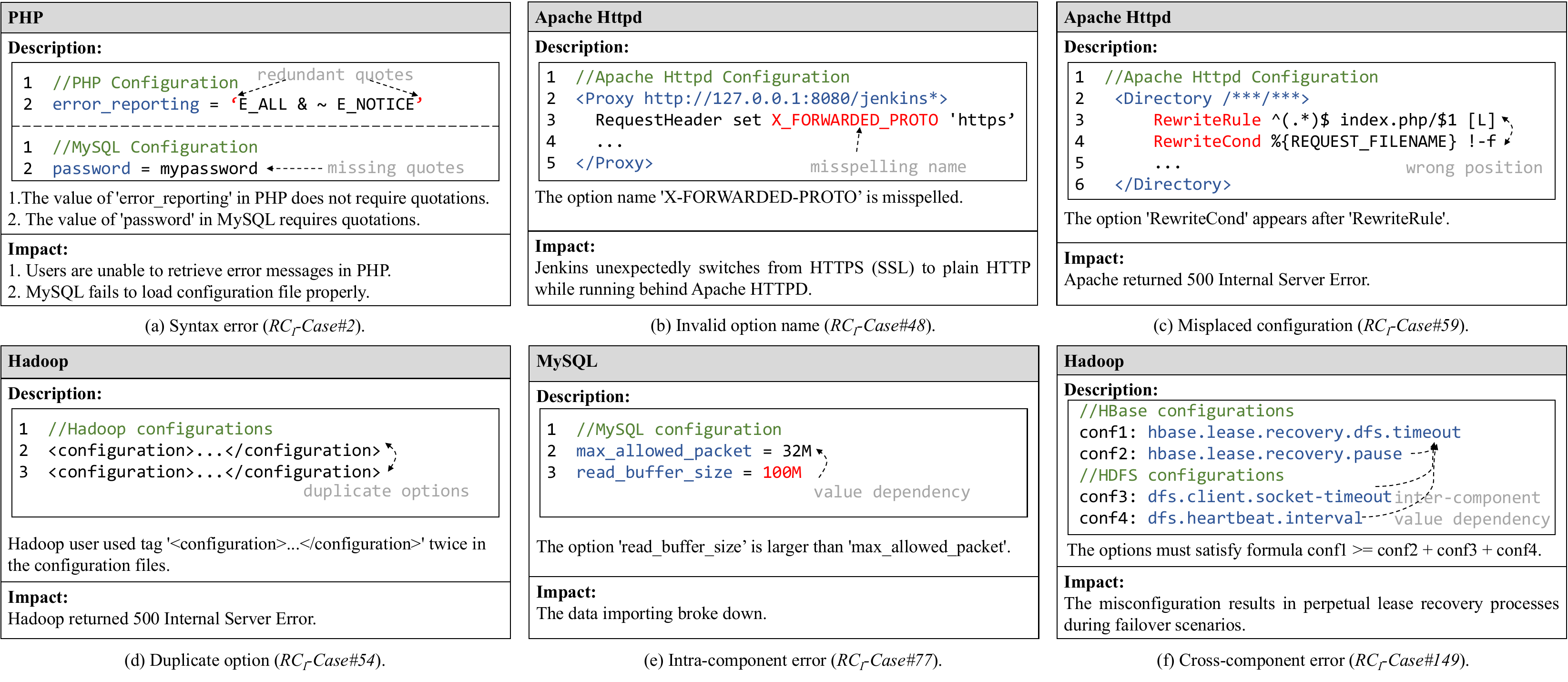}
\caption{Examples of misconfigurations caused by different root causes}
\label{real_cases}
\end{figure*}

Syntax errors occur when the configuration settings violate the way software reads them, e.g., unclosed parentheses, redundant spaces,  unacceptable characters, and symbol misuses. For example, in \textit{RC$\mathit{_{I}}$-Case\#1}, \texttt{query\_cache\_size=128M} was mistakenly written as \texttt{query\_cache\_size-128M}; In \textit{RC$\mathit{_{I}}$-Case\#3}, “=\textgreater” was mistakenly written as “= \textgreater”; In \textit{RC$\mathit{_{I}}$-Case\#4}, the closing bracket was missing. {\textit{Note that we observed 68.52\% (37/54) of syntax errors in our real-world dataset were caused by users' carelessness.}} In addition to carelessness, syntax inconsistency across different platforms and software update also contribute to syntax errors in configuration settings. For instance, as shown in \Cref{real_cases}(a), the setting of the configuration option \texttt{password} in MySQL requires quotation marks (\textit{RC$\mathit{_{I}}$-Case\#35}), while the option \texttt{error\_reporting} in PHP does not use quotation marks (\textit{RC$\mathit{_{I}}$-Case\#2}). Furthermore, users may also be troubled by software update. For example, in \textit{RC$\mathit{_{I}}$-Case\#18}, a \texttt{Grails} user set a configuration value using the syntax of an outdated version.

\subsubsection{Invalid option name}
{Configuration options typically have readable names so that users can understand the role of the configuration. However, we noticed that users often set invalid configuration names. \Cref{real_cases}(b) shows an example of invalid option names, where \texttt{set X-FORWARDED-PROTO} was misspelled as \texttt{set X\_FORWARDED\_PROTO} in \textit{RC$\mathit{_{I}}$-Case\#48}. Since the option \texttt{X-FORWARDED-PROTO} sets the protocol (i.e., HTTP or HTTPS) used by the original client, this misconfiguration led Jenkins to switch from HTTPS to plain HTTP, resulting in potential security risks. In addition, software update may also lead to the changes of option names and thus cause syntax errors when users are not aware of the update. For example, in \textit{RC$\mathit{_{I}}$-Case\#44}, \texttt{autocommit} was an option name in PostgreSQL version 9.5.2, whereas it was removed after PostgreSQL version 9.5. In \textit{RC$\mathit{_{I}}$-Case\#40} and \textit{RC$\mathit{_{I}}$-Case\#43}, users used new features in old versions, leading to invalid option names}.

\subsubsection{Misplaced configuration}
To successfully parse the configurations, users ought to comply with the configuration rules and set the configuration options in the right place. Commonly, modern software often has multiple configuration files that affect different software components. \textit{Among the misplaced configuration issues, 47.62\%~(20/42) of users were confused about the complex file structure of software configurations.} Due to this reason, they may set the configurations in the wrong places, prohibiting the software from correctly parsing the configurations. For example, in \textit{RC$\mathit{_{I}}$-Case\#66}, Apache did not respond to configuration changes because the user modified the configuration file \texttt{httpd.conf}, which was overwritten by the configuration file \texttt{/etc/apache2/sites-enabled/000-default.conf}. Moreover, the order of configuration options may also cause misconfigurations. For example, \Cref{real_cases}(c) shows a software misconfiguration made by a Httpd user (\textit{RC$\mathit{_{I}}$-Case\#59}), where \texttt{RewriteCond} was incorrectly set after \texttt{RewriteRule}, resulting in an internal server error. Finally, sometimes configurations can be divided into different groups based on the scopes of their impacts, so configuration options in different domains have different impacts. For instance, in the configuration file of MySQL, the options on the server and client side are defined by the groups \texttt{[mysqld]} and \texttt{[mysql]}, respectively. However, in \textit{RC$\mathit{_{I}}$-Case\#73}, a MySQL user forgot to specify the group \texttt{[mysqld]} before setting configuration options, leading to a fatal error.

\subsubsection{Duplicate option}
Software users may repeatedly set the same configuration options, resulting in a configuration error. For example, in \Cref{real_cases}(d) (\textit{RC$\mathit{_{I}}$-Case\#54}), a Hadoop user used tag \texttt{<configuration>...</configuration>} twice in four XML files as they had just copied and pasted the configuration files from an online source, raising a fatal error. Another example is that an Nginx user specifies options for the \texttt{listen} directive twice in two configuration files (\textit{RC$\mathit{_{I}}$-Case\#55}). Moreover, the implicit overwriting of software leads users to set redundant options and trigger errors. For example, in Nginx, the default configuration file on \texttt{sites-available} overwrote the user-specified file, {leading to a displaying error (\textit{RC$\mathit{_{I}}$-Case\#106}).}
Note that different software may employ different strategies when parsing repeated configurations. For example, in Nginx, when the option \texttt{try\_files} is specified twice, the second value is overwritten without any errors or warnings. In contrast, the duplicate options in Hadoop caused fatal errors (\textit{RC$\mathit{_{I}}$-Case\#54}).

\subsubsection{Intra-component error}\label{rc1.5}
When there exist multiple configurations in a single software system, users are expected to satisfy the constraints between configurations~(Zhang et al.~\citeyear{2014Encore}). Chen~et al.~(\citeyear{2020cDep}) also highlighted that inter-dependencies among various configuration options significantly contribute to errors in cloud and datacenter systems. \Cref{real_cases}(e) shows an example of intra-component error (\textit{RC$\mathit{_{I}}$-Case\#77}). In MySQL, the option \texttt{read\_buffer\_size} controls the amount of memory allocated for reading rows during a query, while \texttt{max\_allowed\_packet} defines the maximum size of a communication packet. The size of data blocks read during query execution is determined by \texttt{read\_buffer\_size}. In fact, it cannot exceed the value of \texttt{max\_allowed\_packet}. In this case, the value of \texttt{read\_buffer\_size} was set greater than \texttt{max\_allowed\_packet}, causing MySQL to break down during the data import process. However, user manuals often omit such configuration constraints, making it difficult for inexperienced users to correctly configure them.

\subsubsection{Cross-component error}
Components often require some configuration values to meet specific constraints to interact and work properly. Different from intra-component errors, which involve dependencies among configuration options within a single component, cross-component misconfigurations arise when the constraints span across multiple components or software systems. Thus, when dealing with configurations that involve multiple components, modifying a single setting requires making related changes to other settings. As shown in \Cref{real_cases}(f), these four configuration options must satisfy constraint $conf1>=conf2+conf3+conf4$ (\textit{RC$\mathit{_{I}}$-Case\#149}). When the configurations are not properly configured, the lease recovery process will be stuck in endless retries and preemption. Moreover, there are also constraints present within cross-software configurations. In \textit{RC$\mathit{_{I}}$-Case\#164}, there is a mismatch between the Nginx configuration and the Laravel project directory, causing the web server to be unable to locate the requested resources.

\setlength{\parskip}{6pt}
\noindent \textbf{Summary:} According to our dataset, \TODO{171} out of 772 misconfiguration cases were caused by constraint violation. In this paper, we take an in-depth analysis of the constraint violation misconfiguration, and further divide it into six subtypes according to the types of violations. Based on the real-world dataset, we summarize three main reasons that contribute to constraint violation.
\textbf{\ding{172}} Users' carelessness. A majority of syntax errors (i.e., 68.52\%) were caused by users' carelessness, e.g., spelling errors, grammar and punctuation mistakes. \textbf{\ding{173}} Software update. Software updates are often accompanied with newly added configurations, deprecated ones, and even syntax changes of configurations. Configuration errors may occur when users fail to keep pace with such changes and the software lacks backward compatibility {(e.g., \textit{RC$\mathit{_{I}}$-Case\#40}, \textit{RC$\mathit{_{I}}$-Case\#41}, \textit{RC$\mathit{_{I}}$-Case\#43})}. \reviseLiu{\textbf{\ding{174}} Complex configuration relationships. We observe that configurations usually take effect sequentially, which means a configuration might be overwritten by another one if users do not pay attention to the order of configuration settings {(\textit{RC$\mathit{_{I}}$-Case\#75}).} In addition, sometimes there are constraints between multiple configurations within a component. However, these constraints are often ignored by user manuals, thus resulting in intra-component errors {(e.g., \textit{RC$\mathit{_{I}}$-Case\#79}, \textit{RC$\mathit{_{I}}$-Case\#111}).} Finally, the dependencies between multiple components or even software can also result in misconfigurations. For instance, in \textit{RC$\mathit{_{I}}$-Case\#164}, the Nginx configuration did not match the Laravel project directory, leading the web server to be unable to locate the requested resources. }

\subsection{Root Cause 2 (RC$\mathbf{_{II}}$): Resource unavailability}\label{rc2}
{Modern software interacts with various types of resources supplied by the environment, e.g., the file I/O, the network, CPU, and the security mechanisms. It is a common practice that developers provide user-managed options for users to specify their configurations according to the environment. Resource-related configurations must be set correctly to make the necessary resource available.} According to our dataset, resource unavailability accounts for 35.23\% (272/772) of all misconfigurations, and four subtypes of resource unavailability were observed from the dataset. 

\subsubsection{Resource identifier mismatch}
{Software interacts with the resources by specifying the resource identifiers in the configurations. For instance, network resources utilize URIs to denote the protocol schemes, IP addresses, and ports. 
File paths refer to the relative or absolute locations of files. 
However, when resource identifiers point to non-existent resources, resource unavailability misconfiguration occurs. For example, \textit{RC$\mathit{_{II}}$-Case\#88} shows a resource unavailability that Apache cannot log PID to a specified file because it is not placed in the configured path.
Similarly, for online resources, some users mistakenly configured the wrong IP addresses and port numbers. Specifically, in \textit{RC$\mathit{_{II}}$-Case\#67}, the user configured Apache to connect to another software, but the specified port is not listening. Therefore, Apache reported an error of connection.
Moreover, some software allows for URIs with wildcards, regular expressions, and macros. Users must understand the semantics and configure them accurately to match the actual resources. 

\subsubsection{Resource competition}
{Due to limited system resources, software misconfigurations may also lead to resource competition among different software running on the same system. Resource competition occurs when software requires resources that have been occupied and locked by other software. For instance, a user configured Apache's port while a previous configuration had already been set to the same port (\textit{RC$\mathit{_{II}}$-Case\#1}). Hence, Apache failed to start and reported that the address had already been used. According to our dataset, port, device I/O, and database connection are three common types of resources involved in resource competition. In \textit{RC$\mathit{_{II}}$-Case\#2}, the user left many unresponsive Apache processes, which blocked the new Apache instance from accepting the lock of the disk.}

\subsubsection{Unauthorized resource access}
Security policies and mechanisms, such as file access control, network firewall, and TLS certificate, can restrict access to system resources. Systems usually disrupt software behaviors which try to access such protected resources. For example, in \textit{RC$\mathit{_{II}}$-Case\#13}, a user who logged into the system was not authorized to access Apache files, making Apache fail to start. Moreover, for the network firewall, users must specify the particular ports that can be accessed by the software in case the firewall blocks the connection. For instance, in \textit{RC$\mathit{_{II}}$-Case\#41}, the user was unaware of the fact that the firewall policy blocked requests to port 443, resulting in a connection timeout. Finally, if users want to enable HTTPS in their servers, TLS certificates must match the domain name and be registered in advance. Otherwise, certificate misconfigurations may occur (e.g., \textit{RC$\mathit{_{II}}$-Case\#43} and \textit{RC$\mathit{_{II}}$-Case\#19}).

\subsubsection{Hardware limitation}
In addition to the aforementioned resources, hardware is also a significant factor in the software environment. The configuration of a system cannot exceed the hardware limit; otherwise, the software misconfiguration will occur. \textit{RC$\mathit{_{II}}$-Case\#55} is a case where a user configured too much virtual memory for the SVN service, resulting in an out-of-memory issue. Note that some hardware features may require the update of certain firmware or drivers to take effect. For example, in \textit{RC$\mathit{_{II}}$-Case\#56}, a user needed to update the components of the Linux system to support Hyper-V integration.

{\noindent \textbf{Summary: }
According to our dataset, 272 out of 772 misconfiguration cases were caused by resource unavailability, and four subtypes were observed from the dataset. Overall, resource identifier mismatch is the major reason why resources are unavailable when software runs. Specifically, 65.44\% (178/272) of resource unavailability cases occurred because the resource identifiers pointed to non-existent resources. In some cases, although the resources exist, the security policies (e.g., firewall) prohibit software from accessing the resources, accounting for 25.74\% (70/272) of resource issues. Finally, resource competition and hardware limitation may also lead to resource unavailability, where users did not check whether the target resources were idle and enough for the tasks.}

\subsection{Root Cause 3 (RC$\mathbf{_{III}}$): Component integration error}\label{rc3}
{In modern software, users often need to modify configurations to customize the components, including the independent modules and third-party libraries.} {Component integration misconfigurations occur when incorrect deployments of software components are configured.} In our dataset, the component integration errors account for \TODO{12.95\% (100/772)} of all misconfigurations.

\subsubsection{Component incompatibility}
{When the versions of configured components do not match, component incompatibility occurs, accounting for \TODO{58\% (58/100)} of the component integration errors. By analyzing the cases, we find that {56.9\% (33/58)} of component incompatibilities were induced by software upgrade. Users have to upgrade or downgrade the software to the appropriate version. For instance, in \textit{RC$\mathit{_{III}}$-Case\#24}, the Apache failed to run because the version of \texttt{libapache2-mod-php5} did not match Apache v2.4. Moreover, 17 out of {58} component incompatibilities occurred because of the wrong default configurations of components. For example, in \textit{RC$\mathit{_{III}}$-Case\#25}, the TLS protocol configured in Nginx required at least OpenSSL 1.1.1, while the default version of OpenSSL in Amazon Linux was just \texttt{1.0.2k-fips}.

\subsubsection{Component missing}
When necessary components are not correctly configured and thus not identified by software, a configuration error occurs. In most cases, users do not configure the required components inadvertently. Sometimes although users configure the components, they fail to correctly set the locations of components. For example, as shown in \textit{RC$\mathit{_{III}}$-Case\#2}, the user configured Nginx and put the Node.js component in the wrong folder, causing a system failure.

\noindent \textbf{Summary:} \reviseLiu{Component integration errors account for \TODO{12.95\% (100/772)} of the misconfigurations in our dataset. After analyzing the component integration misconfigurations, we categorize them into \reviseLiu{two} types, i.e., component incompatibility and component missing. Specifically, software upgrade is a major reason that induces component incompatibilities, and users need to modify the configurations in time according to the component versions. In addition, necessary components may be missed or incorrectly set (e.g., wrong versions or locations) when users configure software, rendering misconfiguration issues. }

\subsection{Root Cause 4 (RC$\mathbf{_{IV}}$): Configuration Semantic Misinterpretation}\label{rc4}
Configuration options typically employ low-level, implementation-oriented semantics, whereas users focus on high-level functional requirements or observable symptoms. Configuration semantic misinterpretation arises when users fail to map these functional expectations to the corresponding configuration options or values. In our dataset, configuration semantic misinterpretations account for {33.55\% (259/772)} of all misconfiguration cases. \textit{Notably, thirteen issues still remained active after a decade, i.e., still receiving user comments or updates within ten years after the issue was raised.}

\subsubsection{Ambiguous option name}\label{rc4.1}
Ambiguous option names refer to misconfigurations caused by unclear or misleading configuration option names, which prevent users from correctly identifying the option that controls the expected behavior. As a result, users may modify relevant but incorrect options while the effective configuration remains unchanged. For example, in \textit{RC$\mathit{_{IV}}$-Case\#1}, a MySQL user attempted to disable TCP/IP connections by commenting out the port and bind-address options, which led MySQL to fall back to default network settings. The expected behavior can only be achieved by enabling the \texttt{skip-networking} option. The option name does not clearly indicate its role in disabling TCP/IP communication. In addition, the ambiguous option names also confused users when multiple semantically similar options exist across different functionalities or software. In \textit{RC$\mathit{_{IV}}$-Case\#214}, the user confused two similarly named timeout options, \texttt{connectionTimeoutMillis} and \texttt{idleTimeoutMillis}, resulting in idle connections never being released. Moreover, because of different software exposing similarly named configurations, a user increased PHP’s \texttt{upload\_max\_filesize} to allow larger file uploads. However, the actual limit was imposed by Nginx via \texttt{client\_max\_body\_size} (\textit{RC$\mathit{_{IV}}$-Case\#242}).

\subsubsection{Option value and functional requirement mismatch}\label{rc4.2}
Mismatch between option value and functional requirement refers to misconfigurations where users identify the required configuration option, but assign an inappropriate value. The configured values satisfy all syntax and constraint requirements, yet fail to meet the functional requirements of the software. For example, in \textit{RC$\mathit{_{IV}}$-Case\#222}, the user expected to enable SSL functionality and configured SSL certificates in Apache. Due to a lack of understanding of SSL functionality requirements, the user did not enable \texttt{SSLEngine} option in the VirtualHost which is disabled by default. As a result, the Apache server returned an ``ERR\_SSL\_PROTOCOL\_ERROR''. In addition, a user might still set an inappropriate value that satisfied the software parsing specifications after selecting the correct option. Specifically, in \textit{RC$\mathit{_{IV}}$-Case\#4}, a user encountered the error ``\textit{MySQL server has gone away}'' when setting the option \texttt{max\_allowed\_packet} too low. The data import process failed due to internal packet size restrictions, even though the software did not reject the configuration. Such configuring often requires substantial domain knowledge between correct option values and software requirements. Note that {6.31\% (13/206)} of configuration semantic misinterpretation may lead to software performance degradation. In \textit{RC$\mathit{_{IV}}$-Case\#157}, disabling the \texttt{cache} option slows down the data writing to disk.

\noindent \textbf{Summary:} 
Configuration semantic misinterpretation accounts for {33.55\% (259/772)} of the misconfigurations in our dataset. We categorize these issues into two subtypes, i.e., ambiguous option name and mismatch between option value and functional requirement. Specifically, unclear option names or inappropriate values often satisfy syntax constraints but fail to meet functional requirements. Notably, {43.24\% (112/259)} of these misconfigurations result in silent deviations from expected functionalities without obvious symptoms like crashes, often lacking error messages or instructions on how to fix them. Consequently, this root cause is the most prevalent in long-term active issues, accounting for {36.27\% (37/102)} of the cases that remained active for over five years. To mitigate this, providing a checklist of relevant configuration options based on specific error messages or observed symptoms would significantly assist users in troubleshooting.

\begin{keyfindingbox}{Key Findings of RQ1:}
\begin{itemize}[
  label=$\bullet$,
  leftmargin=*,
  itemsep=0.3em
]

    \item To mitigate misconfiguration issues, we summarized the root causes of real-world misconfiguration cases into four groups, i.e., {constraint violation}, {resource unavailability}, {component integration error}, and {configuration semantic misinterpretation}. Each root cause was divided into several subtypes.
    \item A majority of syntax errors (i.e., {68.52\%}) in our dataset were caused by users' carelessness, such as spelling errors, grammar and punctuation mistakes.
    \item Software upgrades are often accompanied with newly added configurations, deprecated ones, changes of configuration syntax and component versions, rendering software misconfigurations like constraint violation and component incompatibilities.
    \item According to our dataset, {65.44\% (178/272)} of resource unavailability cases occurred because the resource identifiers pointed to non-existent resources; the security policies (e.g., firewall) that prohibit software from accessing the resources account for {25.74\% (70/272)} of the resource unavailability issues.
    \item The configuration semantic misinterpretations account for {33.55\% (259/772)} of misconfigurations in our dataset, among which {112} misconfigurations result in silent errors without any error messages or instructions on how to fix them.
    \item In our dataset, 5.02\%~(13/259) of configuration semantic misinterpretations continued to receive attention after a decade, i.e., still receiving user comments or updates within ten years after the issue was raised.

\end{itemize}
\end{keyfindingbox}

\section{RQ2: Trends of Literature}\label{sec5}
In this paper, we also study the trends of literature on software misconfiguration. Specifically, we analyze the literature from five aspects, i.e., misconfiguration symptom, research target, root cause, troubleshooting technique, and debugging artifact.

\begin{table}[htb]
\footnotesize
\centering
\caption{Five types of research targets}
\label{eval_detail}
\begin{tabular}{cc}
\toprule
\textbf{Type} & \textbf{Research Target} \\
\midrule
System software &
GNU core utils, Operating System, etc. \\
Development software & GCC, Jmeter, Randoop, Soot, Jchord, etc.\\
Web application & Internet Explorer, Mozilla, Outlook, etc. \\
Database & Cassandra, MySQL, PostgreSQL, Redis, etc. \\
Cloud software & Azure, Hadoop, OneDrive, Openstack, etc. \\
\bottomrule
\end{tabular}
\end{table}

\subsection{Research targets and misconfiguration symptoms}\label{sec5:1}
We first study the trends in research targets and the symptoms of misconfiguration that attract the attention of researchers in the field of misconfiguration troubleshooting. {To investigate the research targets, we categorize them into five
groups,} i.e., system software, development software, web application, database, and cloud software. \Cref{eval_detail} exemplifies the instances of each group.

\subsubsection{Trends of research targets}\label{sec5:1:1}
\Cref{evolution_sec_a}(a) shows the changes of research targets in the field of misconfiguration troubleshooting from 2003 to 2024. First, we find that researchers had been interested in misconfigurations occurring in system software until 2012. After 2012, few studies were conducted on system software. In contrast, the number of web applications as the research targets has a stable growth during the past 22 years. By analyzing the collected papers, we find that for web applications, researchers first focused on browsers (e.g., Internet Explorer) and email systems (e.g., Outlook), and then turned their eyes to web servers (e.g., {Apache Httpd}). The trends of system software and web applications reveal the significance of {infrastructure} software in the research field at early years. In recent years, except web applications, database and cloud software have also attracted increasing research attention, due to the rapid growing demand for storage and cloud service.

\begin{figure}[htb]
\centering  
\subfigure[Research targets]{ 
\begin{minipage}{0.45\hsize}
\centering    
\includegraphics[scale=0.18]{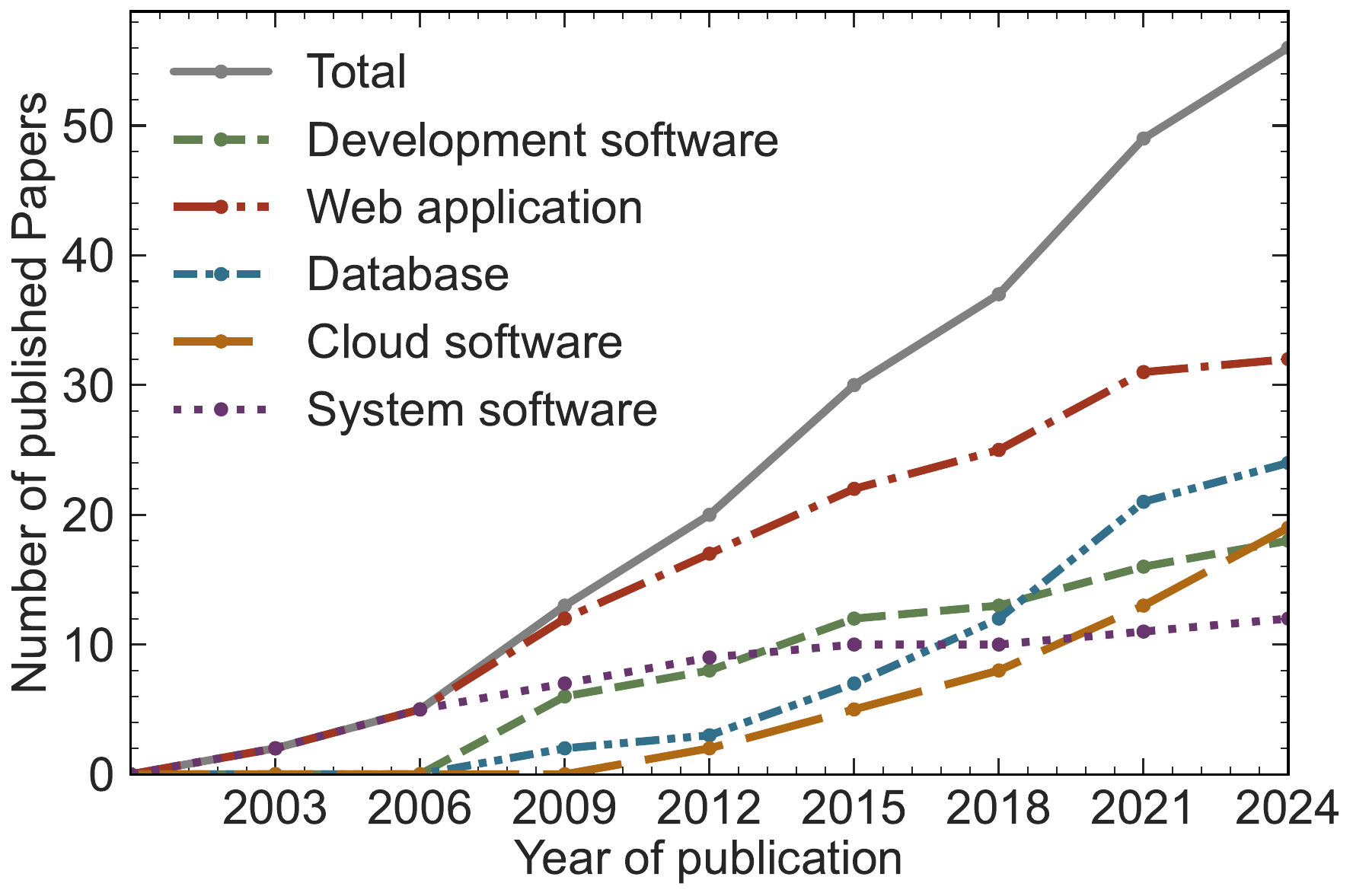} \label{fig:5_1}
\end{minipage}
}
\subfigure[Misconfiguration symptoms]{ 
\begin{minipage}{0.45\hsize}
\centering    
\includegraphics[scale=0.18]{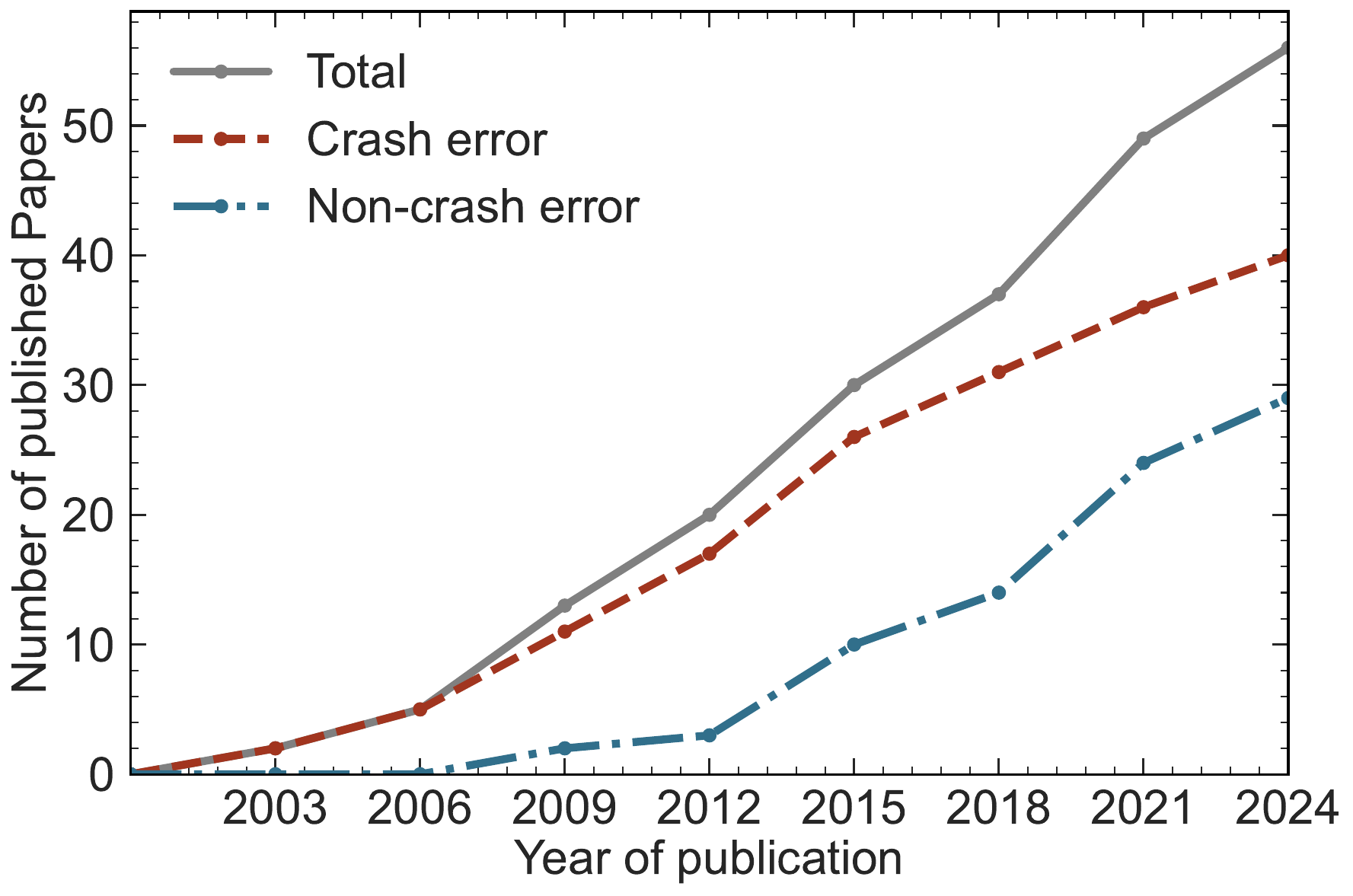} \label{fig:5_2}
\end{minipage}
}

\caption{Trends of research targets and misconfiguration symptoms from 2003 to 2024}    
\label{evolution_sec_a}    
\end{figure}

\subsubsection{Trends of misconfiguration symptoms} 
To investigate the symptoms of software misconfigurations, we classify the symptoms into two types, i.e., crash and non-crash errors, as shown in \Cref{statistics-study}. Crash errors refer to software failures that can be easily observed, and non-crash errors refer to software behaviors deviated from expectations (e.g., performance degradation). \Cref{evolution_sec_a}(b) shows the changes of misconfiguration symptoms addressed by previous studies. It can be seen that almost all studies before 2012 concentrated on crash errors caused by misconfigurations. For example, Whitaker~et al.~(\citeyear{2004Chronus}) studied the configuration bugs that made the forward and backward buttons in Mozilla invalid. {After 2012,} the research on non-crash misconfigurations has a significant growth. These observations of misconfiguration symptoms in the literature are consistent with the trends of research targets, since users have different expectations for different software. For instance, database users not only expect the database to perform correctly, but also have requirements for database performance. Therefore, it is desirable to address with misconfigurations that lead to performance degradation of database systems~(Mahgoub et al.~\citeyear{mahgoub2019sophia}; Li et al.~\citeyear{li2020learnconf}).

\noindent\textbf{Summary: }The research targets and misconfiguration symptoms have important changes over the past two decades. Specifically, \textbf{\ding{172}} the research targets have changed from system and {infrastructure} software (e.g., Outlook) to advanced applications, including web applications, cloud software, and databases, reflecting the rapid growing demand for storage and cloud service. \textbf{\ding{173}} Early studies towards misconfiguration troubleshooting concentrated on crash errors. However, after 2012, the research on non-crash errors such as performance degradation has a significant growth. \textbf{\ding{174}} The trends of misconfiguration consequences in the literature are consistent with the trends of research targets, {i.e., from crash errors in system and infrastructure software to non-crash errors in advanced applications.} 

\subsection{Troubleshooting techniques and debugging artifacts}\label{sec5:2}

We also investigate the evolution of techniques of misconfiguration troubleshooting, along with the debugging artifacts used in the techniques.

\subsubsection{Evolution of troubleshooting techniques}
\Cref{evolution_sec_b}(a) illustrates the evolution of misconfiguration troubleshooting techniques from 2003 to 2024. We classify the techniques into five groups, i.e., the statistic-based methods, deterministic replay, program analysis, and the learning-based methods. From \Cref{evolution_sec_b}(a), it can be seen that a majority of studies before 2009 utilized the statistic-based methods to address misconfiguration issues. However, after 2012, only a few studies employed the statistic-based methods and deterministic replay on misconfiguration troubleshooting. Instead, program analysis and the learning-based methods become more popular in recent years, especially machine learning (e.g., large language model) has demonstrated their capabilities in modeling patterns of large-scale systems, making it a promising approach for misconfiguration troubleshooting. 

\begin{figure}[htbp]
\centering  
\subfigure[Troubleshooting techniques]{ 
\begin{minipage}{0.45\hsize}
\centering    
\includegraphics[scale=0.18]{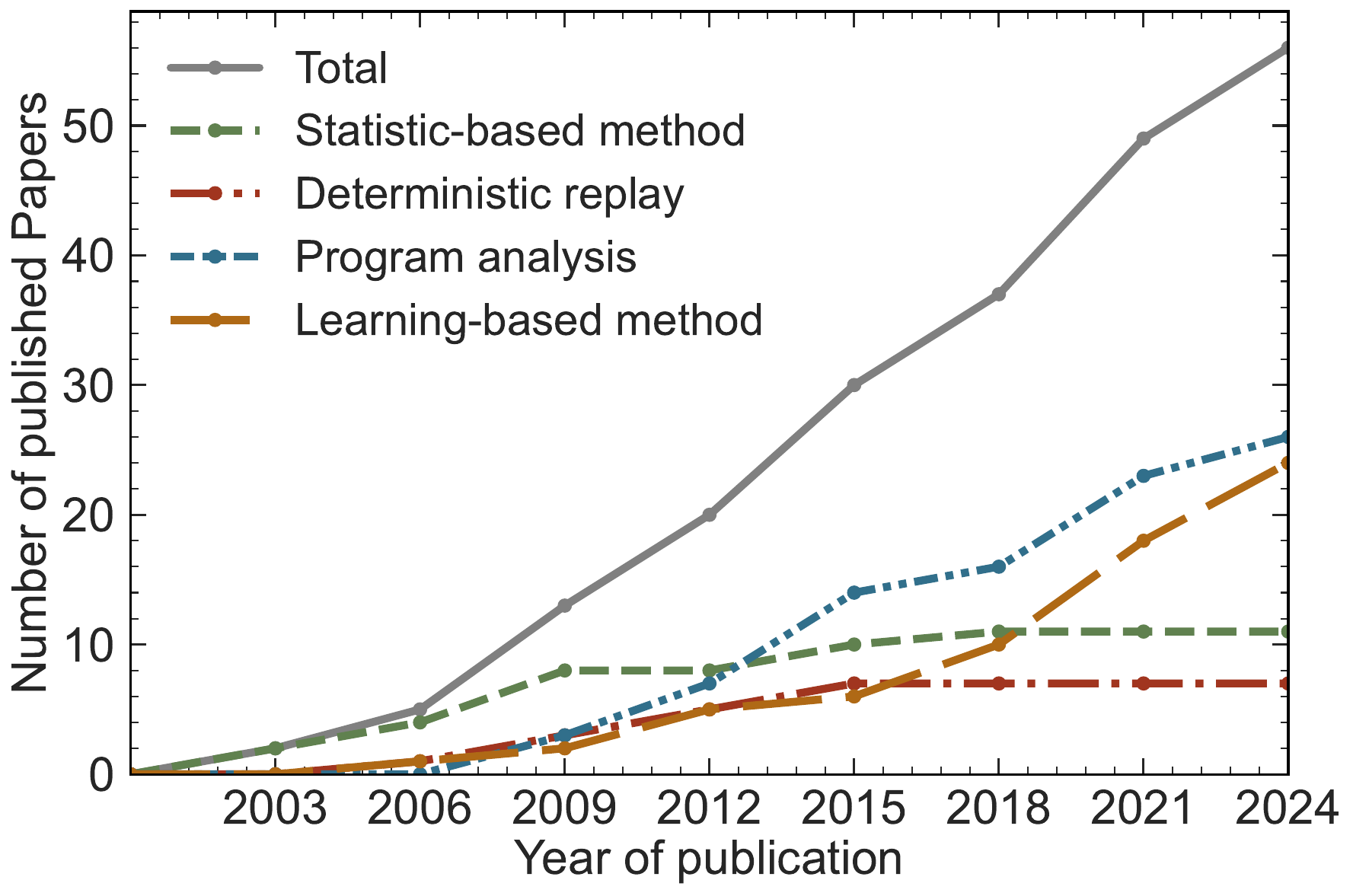}\label{fig:5_3}
\end{minipage}
}
\subfigure[Debugging artifacts]{ 
\begin{minipage}{0.45\hsize}
\centering    
\includegraphics[scale=0.18]{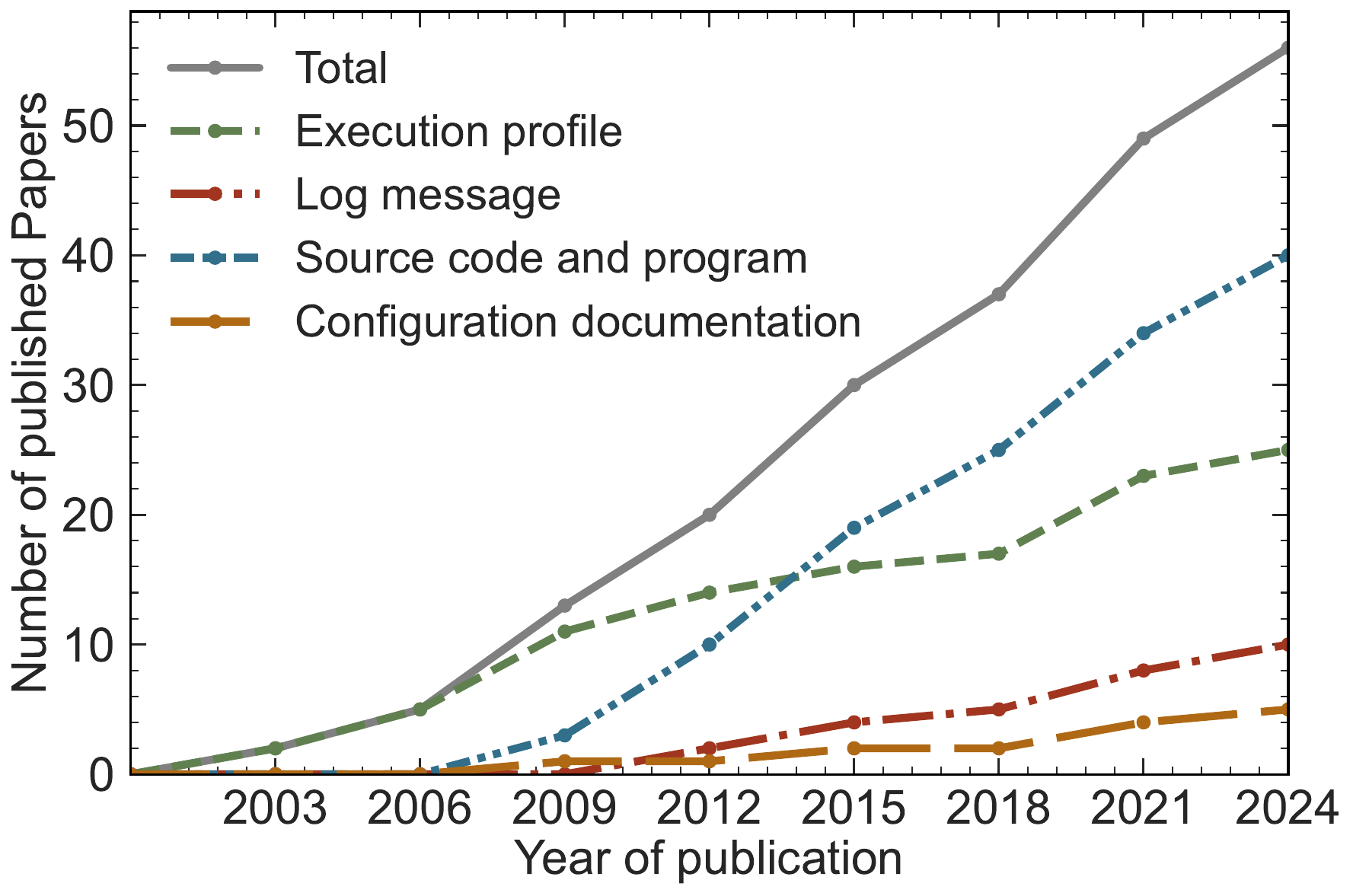}\label{fig:5_4}
\end{minipage}
}

\caption{Evolution of troubleshooting techniques and debugging artifacts from 2003 to 2024}    
\label{evolution_sec_b}    
\end{figure}

\subsubsection{Changes of debugging artifact}
{To investigate the debugging artifacts used by researchers to address misconfigurations, we categorize the debugging artifacts employed in the literature into four types, i.e., execution profile, log message, source code and program, and documents. \Cref{evolution_sec_b}(b) shows the changes of debugging artifacts from 2003 to 2024. First, it can be seen that the execution profiles have been widely used for debugging misconfigurations. The reason is that both the statistic-based methods and the learning-based methods usually rely on execution profiles to diagnose misconfigurations. Moreover, it can also be observed that the trend of source code and program is consistent with the trend of program analysis in \Cref{evolution_sec_b}(a). Finally, documents (e.g., user manuals) and log messages also increase gradually after 2018. As the natural language processing (i.e., NLP) techniques have achieved great success in many research fields, some researchers applied NLP techniques to extract the constraints and patterns in user manuals and log messages. }

{\noindent \textbf{Summary: } {Over the past two decades,} the techniques and artifacts used for misconfiguration troubleshooting have evolved simultaneously. {The detailed information of each paper can be seen in \Cref{statistics-study}.} Specifically, \textbf{\ding{172}} although widely used in the early years, the statistic-based methods and deterministic replay had been abandoned in the field of misconfiguration troubleshooting after 2012. \textbf{\ding{173}} Program analysis and the learning-based methods are the mainstream methods in recent years, which promote the use of source code, programs, and execution profiles as the debugging artifacts.
\textbf{\ding{174}} With the successful applications of the NLP techniques in many fields, they have gained more visibility for their capabilities to analyze configuration documents and log messages, so as to assist misconfiguration troubleshooting.}

\begin{figure}[htb]
\centering  
\subfigure[Root causes]{ 
\begin{minipage}{0.45\hsize}
\centering    
\includegraphics[scale=0.18]{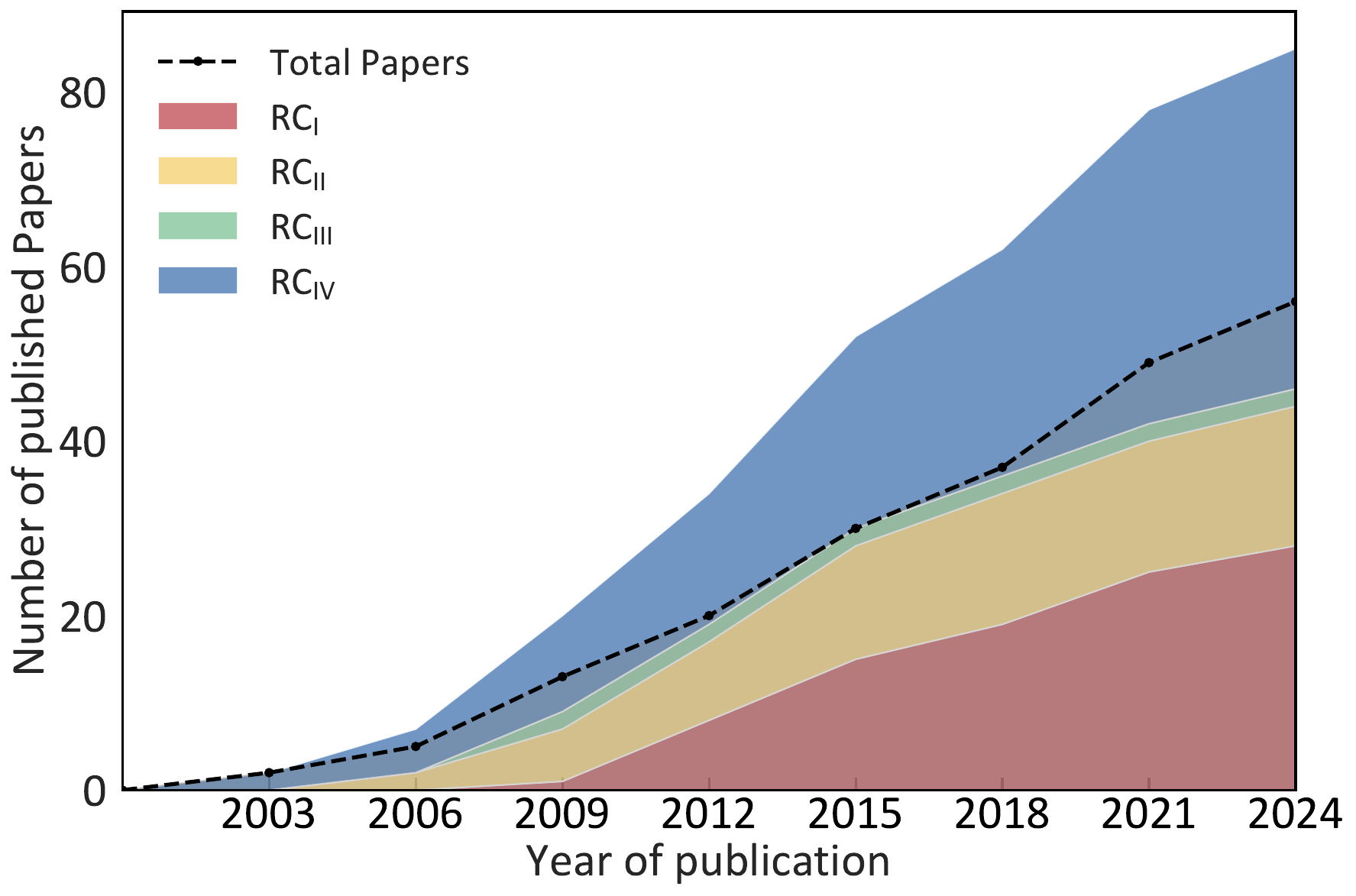} \label{fig:5_5}
\end{minipage}
}
\subfigure[Root cause subtypes]{ 
\begin{minipage}{0.45\hsize}
\centering    
\includegraphics[scale=0.18]{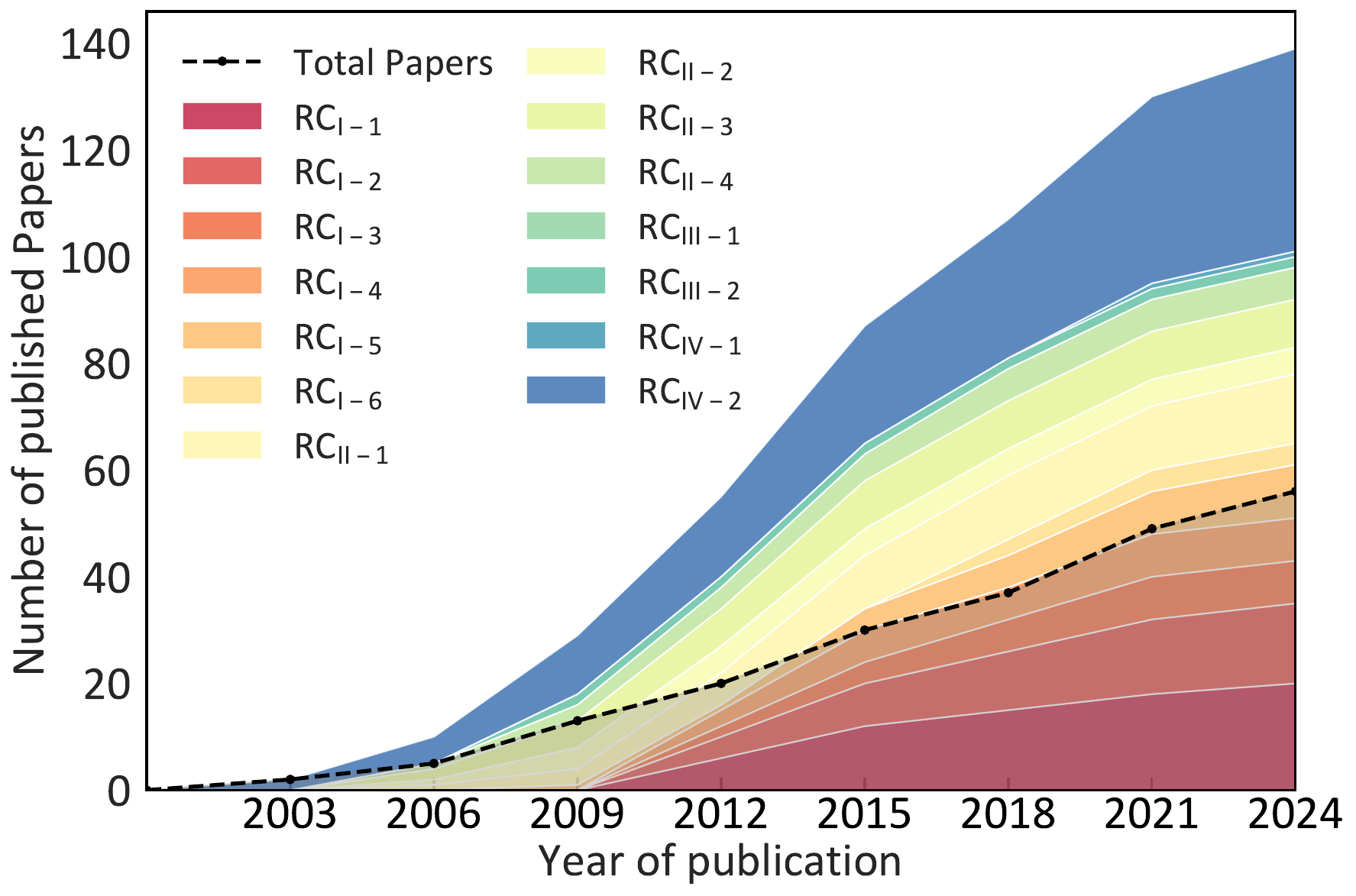} \label{fig:5_6}
\end{minipage}
}

\caption{Trends of the addressed root causes and its subtypes from 2003 to 2024}    
\label{evolution_sec_c}    
\end{figure}

\subsection{Trends of addressed root causes} 
To investigate the root causes addressed by existing troubleshooting techniques, we categorized the collected papers according to the root causes that they could address. \Cref{evolution_sec_c} illustrates the evolutionary trends of the root causes targeted by the literature from 2003 to 2024. First, as shown in \Cref{evolution_sec_c}(a), configuration semantic misinterpretations have been the most frequently studied category over the past two decades. This aligns with our RQ1 finding that this type of misconfigurations is a major and long-lasting challenge for users. However, within this category, we observe that few papers address misconfigurations caused by ambiguous option names in \Cref{evolution_sec_c}(b). \reviseLiu{Second, component integration errors receive the least attention from the research community. Studies related to component integration errors only show a slight increase after 2006 and remain much fewer than those on constraint violation, resource unavailability, and configuration semantic misinterpretation. Finally, the research on constraint violations experienced a noticeable increase starting from 2009. Researchers paid little attention to syntax or constraint errors in the early years. After 2015, studies on multi-option errors increased substantially.}

\begin{keyfindingbox}{Key Findings of RQ2:}
\begin{itemize}[
  label=$\bullet$,
  leftmargin=*,
  itemsep=0.3em
]

    \item Over the past two decades, the research targets have changed from system and infrastructure software (e.g., Outlook) to advanced applications (e.g., cloud service). 
    \item Early studies focused on misconfigurations that led to obvious consequences (e.g., crashes), while after 2012, the research on non-crash errors such as performance degradation and security risks has a significant growth.
    \item The statistic-based methods and deterministic replay had been abandoned in the field of misconfiguration troubleshooting after 2012 though they were widely used in the early years.
    \item Program analysis and the learning-based methods are the mainstream techniques for misconfiguration troubleshooting, and the NLP techniques have gained more visibility for their capabilities to analyze configuration documents and logs. 
    \item Configuration semantic misinterpretations have consistently dominated the literature, yet very few studies address misconfigurations caused by ambiguous option names.
    \item Component integration errors receive the least attention, and research on constraint violations has steadily increased since 2009.
    
\end{itemize}
\end{keyfindingbox}

\begin{table*}[htb]
\footnotesize
\centering
\caption{The statistics of misconfiguration troubleshooting studies}
\label{statistics-study}
\resizebox{0.96\hsize}{!}{
\renewcommand{\arraystretch}{1.3}
\begin{adjustbox}{width=\textwidth} 
    \begin{tabular}{ccccccccc}
    \toprule
    \multirow{2}{*}{\textbf{Literature}} & \multirow{2}{*}{\textbf{Sympt.}} & \multirow{2}{*}{\textbf{Target}} &  \multirow{2}{*}{\textbf{Tech.}} & \multirow{2}{*}{\textbf{Artfct.}} & \multicolumn{3}{c}{\textbf{Tool}} & \multirow{1}{*}{\textbf{Dataset}} \\
    \cline{6-8}
    &&&&& \textbf{Avail.} & \textbf{Maint.} & \textbf{Adapt.} & \textbf{Avail.}  \\
    \midrule

    PeerPressure~(Wang et al.~\citeyear{2003peerpressure,2004peerpressure-osdi}) & Crash & SS, WA & Stat. & EP & \Circle & \Circle & \Circle & \Circle \\
    Strider~(Wang et al.~\citeyear{2004strider}) & Crash & SS, WA & Stat. & EP & \Circle & \Circle & \Circle & \Circle \\
    Chronus~(Whitaker et al.~\citeyear{2004Chronus}) & Crash & SS, WA & Stat., Rep. & EP & \Circle & \Circle & \Circle & \Circle \\
    Lao et al.~(\citeyear{2004state-based}) & Crash & SS, WA & Stat. & EP & \Circle & \Circle & \Circle & \Circle \\
    Yuan et al.~(\citeyear{2006EventTraceBased}) & Crash & SS, WA & Learn. & EP & \Circle & \Circle & \Circle & \Circle \\
    Triage~(Tucek et al.~\citeyear{tucek2007triage}) & Crash & SS, DS, WA, DB & Rep., PA & EP, Src. & \Circle & \Circle & \Circle & \Circle \\
    Autobash~(Su et al.~\citeyear{2007autobash}) & Crash & DS, WA & Stat., Rep. & EP & \Circle & \Circle & \Circle & \Circle \\
    Snitch~(Mickens et al.~\citeyear{2007snitch}) & Crash & WA & Learn. & EP & \Circle & \Circle & \Circle & \Circle \\
    SigConf~(Attariyan and Flinn~\citeyear{2008sigConf}) & Crash & DS, WA & Stat. & EP & \Circle & \Circle & \Circle & \Circle \\
    Ding et al.~(\citeyear{2008signature-based}) & Crash & DS, WA, DB & Stat. & EP & \Circle & \Circle & \Circle & \Circle \\
    ConfigRE~(Wang et al.~\citeyear{2008configre}) & Non-crash & WA & PA & Src. & \Circle & \Circle & \Circle & \Circle \\
    CURE~(White et al.~\citeyear{2008CURE,2010CURE-JSS}) & Non-crash & SS, DS & PA & Src., Doc. & \Circle & \Circle & \Circle & \Circle \\
    Su and Flinn~(\citeyear{2009predicate-based}) & Crash & DS, WA & Stat. & EP & \Circle & \Circle & \Circle & \Circle \\
    ConfAid~(Attariyan and Flinn~\citeyear{2010confAid}) & Crash & WA & Rep., PA & EP, Src. & \Circle & \Circle & \Circle & \Circle \\
    SherLog~(Yuan et al.~\citeyear{2010sherlog}) & Crash & SS, DS, WA & PA & Log., Src. & \Circle & \Circle & \Circle & \Circle \\
    ConfAnalyzer~(Rabkin and Katz~\citeyear{2011ConfAnalyzer,2011ConfAnalyzer_issta}) & Crash & DS, CL & PA & Log., Src. & \CIRCLE & \Circle & \Circle & \Circle \\
    CODE~(Yuan et al.~\citeyear{2011CODE}) & Crash & SS, WA & Learn. & EP, Src. & \Circle & \Circle & \Circle & \Circle \\
    Wang et al.~(\citeyear{2011machineLear-based}) & Crash & WA & Learn. & Src. & \Circle & \Circle & \Circle & \Circle \\
    X-ray~(Attariyan et al.~\citeyear{2012Xray}) & Non-crash & WA, DB & Rep., PA & EP, Src. & \Circle & \Circle & \Circle & \Circle \\
    Uchiumi et al.~(\citeyear{2012Cloud-basedMisDetec}) & Crash & CL & Learn. & Src. & \Circle & \Circle & \Circle & \Circle \\
    SPEX~(Xu et al.~\citeyear{2013SPEX}) & Crash, Non-crash & WA, DB & PA & Src. & \Circle & \Circle & \Circle & \Circle \\
    ConfDiagnoser~(Zhang and Ernst~\citeyear{2013confDiagnoser}, \citeyear{2013confdiagnoser_3page}) & Crash, Non-crash & DS & Rep., PA & EP, Src. & \CIRCLE & \Circle & \Circle & \Circle \\
    ConfDebugger~(Dong et al.~\citeyear{2013confDebugger}) & Crash & DS & PA & Log., Src. & \Circle & \Circle & \Circle & \Circle \\
    Kannan and Bhamidipaty~(\citeyear{2013Cloud-based}) & Crash & WA, DB, CL & Stat. & Src. & \Circle & \Circle & \Circle & \Circle \\
    Confeagle~(Eshete et al.~\citeyear{2013confeagle}) & Non-crash & WA & PA & Src. & \Circle & \Circle & \Circle & \Circle \\
    SmartFixer~(Wang et al.~\citeyear{2013SmartFixer}) & Crash, Non-crash & SS & PA & Doc. & \CIRCLE & \Circle & \Circle & \Circle \\
    ConfSuggester~(Zhang and Ernst~\citeyear{2014confSuggester}) & Crash, Non-crash & DS, WA, DB & Rep., PA & EP, Src. & \CIRCLE & \Circle & \Circle & \Circle \\
    Encore~(Zhang et al.~\citeyear{2014Encore}) & Crash, Non-crash & WA, DB & Learn. & Src. & \Circle & \Circle & \Circle & \Circle \\
    ConfDoctor~(Dong et al.~\citeyear{2015confDoctor}) & Crash & DS, CL & PA & Log., Src. & \Circle & \Circle & \Circle & \Circle \\
    ConfValley~(Huang et al.~\citeyear{2015confvalley}) & Crash, Non-crash & CL & Stat. & Src. & \Circle & \Circle & \Circle & \Circle \\
    PCHECK~(Xu et al.~\citeyear{2016PCHECK}) & Crash & WA, DB, CL & PA & Src. & \Circle & \Circle & \Circle & \LEFTcircle \\
    ConfigC~(Santolucito et al.~\citeyear{2016configc}) & Crash, Non-crash & DB & Learn. & Src. & \Circle & \Circle & \Circle & \Circle \\
    CsCEs~(Sayagh et al.~\citeyear{2017Cross-stack}) & Crash & DS, WA, DB & PA & Src. & \Circle & \Circle & \Circle & \Circle \\
    ConfigV~(Santolucito et al.~\citeyear{2017configv}) & Crash, Non-crash & DB & Learn. & Src. & \CIRCLE & \CIRCLE & \CIRCLE & \CIRCLE \\
    Dexter~(Talwadker~\citeyear{2017dexter}) & Crash & WA & Stat. & Log. & \Circle & \Circle & \Circle & \Circle \\
    SmartConf~(Wang et al.~\citeyear{wang2018smartconf}) & Non-crash & DB, CL & Learn. & EP, Src. & \Circle & \Circle & \Circle & \Circle \\
    Scout~(Hsu et al.~\citeyear{hsu2018scout}) & Non-crash & CL & Learn. & Src. & \Circle & \Circle & \Circle & \Circle \\
    Sophia~(Mahgoub et al.~\citeyear{mahgoub2019sophia}) & Non-crash & DB & PA, Learn. & EP, Src. & \Circle & \Circle & \Circle & \Circle \\
    Configcrusher~(Velez et al.~\citeyear{2020configcrusher}) & Non-crash & DS & PA & EP, Src. & \CIRCLE & \Circle & \Circle & \Circle \\
    Rex~(Mehta et al.~\citeyear{2020rex}) & Non-crash & CL & PA, Learn. & Log., Src. & \Circle & \Circle & \Circle & \Circle \\
    PracExtractor~(Xiang et al.~\citeyear{2020pracextractor}) & Crash, Non-crash & WA, DB, CL & Learn. & Doc. & \Circle & \Circle & \Circle & \CIRCLE \\
    Robert et al.~(\citeyear{2020machineLear-based3}) & Crash & CL & Learn. & EP, Log., Src. & \Circle & \Circle & \Circle & \Circle \\
    LearnConf~(Li et al.~\citeyear{li2020learnconf}) & Non-crash & DB, CL & PA & Src. & \Circle & \Circle & \Circle & \Circle \\
    Violet (Hu et al.~\citeyear{2020Violet}) & Non-crash & DB, WA & PA & EP, Src. & \CIRCLE & \Circle & \Circle & \Circle \\
    ConfigX~(Zhang et al.~\citeyear{zhang2021configX}) & Crash, Non-crash & WA, DB & PA & Src. & \Circle & \Circle & \Circle & \LEFTcircle \\
    ConfProf~(Han et al.~\citeyear{han2021confprof}) & Non-crash & DS, WA, DB & PA, Learn. & EP, Src. & \Circle & \Circle & \Circle & \Circle \\
    ConfDetect~(Li et al.~\citeyear{2021ConfDetect}) & Crash & WA, DB & Learn. & Log., Src. & \Circle & \Circle & \Circle & \Circle \\
    ConfigMiner~(Sayagh et al.~\citeyear{2021ConfigMiner}) & Crash, Non-crash & WA, DB, CL & Learn. & Doc. & \Circle & \Circle & \Circle & \Circle \\
    Weber et al.~(\citeyear{2021white-box}) & Non-crash & SS, DS, DB & Learn. & EP & \CIRCLE & \Circle & \Circle & \Circle \\
    AgileCtrl~(Wang et al.~\citeyear{wang2022agilectrl}) & Non-crash & DB, CL & Learn. & EP, Src. & \Circle & \Circle & \Circle & \Circle \\
    SafeTune~(He et al.~\citeyear{he2022safetune}) & Non-crash & WA, DB, CL & Learn. & Src., Doc. & \CIRCLE & \CIRCLE & \CIRCLE & \Circle \\
    Xu et al.~(\citeyear{ding2023real}) & Crash & CL & PA, Learn. & Log., Src. & \Circle & \Circle & \Circle & \Circle \\
    MMD~(Zhou et al.~\citeyear{2023MMD}) & Crash, Non-crash & DS, CL & PA & Src. & \Circle & \Circle & \Circle & \CIRCLE \\
    DiagConfig~(Chen et al.~\citeyear{2023DiagConfig}) & Non-crash & SS, DS, DB & PA, Learn. & EP, Src. & \Circle & \Circle & \Circle & \Circle \\
    LogConfigLocalizer~(Shan et al.~\citeyear{2024LogConfigLocalizer}) & Crash & CL & Learn. & Log. & \CIRCLE & \CIRCLE & \Circle & \Circle \\
    SlsDetector~(Wen et al.~\citeyear{2024SlsDetector}) & Crash, Non-crash & CL & Learn. & Src. & \CIRCLE & \CIRCLE & \Circle & \CIRCLE \\
    \bottomrule
    \end{tabular}
    \end{adjustbox}
}
\begin{tablenotes}
\tiny
\item[] Sympt. (Abbr. for Software symptom): \textit{Crash} denotes crash error, and \textit{Non-crash} denotes non-crash error;\\
Target (Abbr. for Research target): \textit{SS} denotes system software, \textit{DS} denotes development software, \textit{WA} denotes web application, \textit{DB} denotes database, and \textit{CL} denotes cloud software;\\
Tech. (Abbr. for Troubleshooting technique): \textit{Stat.} denotes statistic-based methods, \textit{Rep.} denotes deterministic replay, \textit{PA} denotes program analysis, \textit{Learn.} denotes learning-based methods;\\
Artfct. (Abbr. for Debugging artifact): \textit{EP} denotes execution profile, \textit{Log.} denotes log message, \textit{Src.} denotes source code and program, \textit{Doc.} denotes documents;\\
Avail. (Abbr. for Availability): \CIRCLE~ means the tool is publicly available, and \Circle~ means the tool is not available; \\
Maint. (Abbr. for Maintenance): \CIRCLE~ means the project has been updated in the last three years, and \Circle~ means it has not been updated in the last three years; \\
Adapt. (Abbr. for Adaptability): \CIRCLE~ means the tool can be adapted to other research targets, and \Circle~ means the adaptability of the tool is not good; \\
Dataset Avail. (Abbr. for Dataset Availability): \CIRCLE~ means the dataset used by the paper is publicly available, \LEFTcircle, means the dataset is partly available, \Circle~ means not available.
\end{tablenotes}
\end{table*}

\section{RQ3: Available Tools and Datasets}\label{sec6}
In this section, we conduct a comprehensive study on the availability, maintainability, and adaptability of misconfiguration troubleshooting tools and the availability of misconfiguration datasets. {\textit{Note that we would like to conduct an empirical study on the effectiveness of existing tools; however, the unavailability of most existing tools and datasets makes it difficult to conduct comprehensive and fair experiments for tool evaluation in this field.}}

\subsection{Misconfiguration troubleshooting tools}\label{sec6:1}
To investigate practicality of existing tools for misconfiguration troubleshooting, first we tried to acquire all the tools presented in the published papers, and then we searched with the tool names on developer platforms (i.e., GitHub, Bitbucket, and Google Code). However, \textit{only \TODO{11} out of \TODO{60} papers {{(i.e., ConfAnalyzer\footnote{\url{https://github.com/asrabkin/Confalyzer/}}, ConfDiagnoser\footnote{\url{https://code.google.com/archive/p/config-errors/}}, ConfSuggester\footnotemark[\value{footnote}], SmartFixer\footnote{\url{https://code.google.com/archive/p/smart-fixer/}}, ConfigV\footnote{\url{https://github.com/ConfigV/ConfigV/}}, ConfigCrusher\footnote{\url{https://github.com/miguelvelezmj25/asej20-sm/}}, Violet\footnote{\url{https://github.com/OrderLab/violet/}}, white-box tool\footnote{\url{https://github.com/AI-4-SE/White-Box-Performance-Influence-Models/}}, SafeTune\footnote{\url{https://github.com/TimHe95/SafeTune/}}, LogConfigLocalizer\footnote{\url{https://github.com/shanshw/LogConfigLocalizer/}} and SlsDetector\footnote{\url{https://github.com/WenJinfeng/SlsDetector_ConfigurationDetection/}}) were publicly available until December 2024, among which only ConfigV, SafeTune, LogConfigLocalizer and SlsDetector have updates in the last three years}}}. 

In addition, we also investigate the adaptability of the tools to the research targets not presented in their papers.
First, we observe that \TODO{9} out of \TODO{11} available tools {(i.e., ConfAnalyzer, ConfDiagnoser, ConfSuggester, SmartFixer, ConfigCrusher, Violet, white-box tool, LogConfigLocalizer, and SlsDetector)} were implemented in conjunction with the research targets they used, leading to poor adaptability to other research targets. For example, configuration options were manually designated and hard coded in their tool implementations.
Second, the environmental requirements of the tools also limit the use of these tools. {For example, ConfDiagnoser and ConfSuggester exploited WALA 1.3.4\footnote{\url{https://github.com/wala/WALA/}} in their tools, which was only supported by some outdated JDK versions (e.g., JDK 1.6). Hence, these tools should be updated so that users can directly use these tools in their environments.} 
Third, these tools were implemented for specific programming languages, which also limits their adaptability to software written in other programming languages. 
{For example, Java-based misconfiguration troubleshooting tools~(Rabkin and Katz~\citeyear{2011ConfAnalyzer}; 
Zhang and Ernst~\citeyear{2013confDiagnoser}, \citeyear{2014confSuggester}) typically use specialized Java-based program analysis frameworks, i.e, Soot~(Vallée-Rai et al.~\citeyear{soot}) and WALA\footnote{\url{http://wala.sf.net/}}, which cannot be directly applied to software written in C/C++.}

{\noindent \textbf{Summary: } 
We tried our best to obtain troubleshooting tools for misconfiguration. However, only \TODO{11} out of {60} papers made their tools publicly available; only {4} tools have been updated in the past three years. Even worse, among the available tools, \TODO{9} tools were implemented tightly coupled with their research targets in the papers, rendering the poor adaptability of these tools. In addition, the environmental requirements (e.g., an outdated version JDK 1.6) also limit the use of these tools.}

\subsection{Misconfiguration datasets}\label{sec6:2}

To investigate the quality of the datasets for misconfiguration troubleshooting, we collected the datasets from previous studies and compared them with the real-world dataset we collected in this paper. In total, we find that \textit{only {6} out of \TODO{60} studies used publicly available datasets, among which 4 studies used the same dataset~(Xu et al.~\citeyear{xu2015knobs}), one study only focused on multi-option errors~(Zhou et al.~\citeyear{2023MMD}), and another study used the dataset with crafted errors~(Wen et al.~\citeyear{2024SlsDetector}).}\footnote{The experiments were conducted in December 2024.}  

\begin{table}[htb]
\footnotesize
\centering
\caption{Comparison between our dataset and available datasets}
\label{dataset_cmp}
\begin{tabular}{ccccc}
\toprule
\multirow{1}{*}{\textbf{Dataset}} & \textbf{Real-world} & \textbf{Root cause} & \textbf{\#Errors} & \textbf{\#Software} \\
\midrule
Xu~et al.~(\citeyear{misdataset-xu}) &
\CIRCLE & \Circle & 98 & 3 \\
MMD~(Zhou et al.~\citeyear{2023MMD}) &
\LEFTcircle & \Circle & 22 & 4 \\
SlsDetector~(Wen et al.~\citeyear{misdataset-wen}) & \LEFTcircle & \CIRCLE & 84 & 1 \\
This work &
\CIRCLE & \CIRCLE & {772} & 7 \\
\bottomrule
\end{tabular}
\begin{tablenotes}\centering
\scriptsize
\item[] Real-world: \CIRCLE~ means the dataset only include real-world errors, and \LEFTcircle~ means the dataset include real-world and manually crafted errors; \\
Root cause: \CIRCLE~ means the dataset is labeled with the root causes of misconfigurations, and \Circle~ means the dataset is not labeled.
\end{tablenotes}
\end{table}

{\Cref{dataset_cmp} shows the statistics of the datasets. First, it can be seen that all three datasets contain real-world misconfiguration issues. Specifically, one dataset contains 98 real-world configuration errors for three research targets~(Xu~et al.~\citeyear{misdataset-xu}). The dataset of MMD~(Zhou et al.~\citeyear{2023MMD}) contains 14 real-world misconfigurations and 8 manually crafted ones for four research targets. The dataset of SlsDetector~(Wen et al.~\citeyear{2024SlsDetector}) includes 26 injected configuration errors and 58 real-world misconfigurations collected from Github issues. In this work, we collected a dataset comprising {772} real-world configuration issues on 7 research targets. 
Second, only one existing dataset provides root cause labels for misconfigurations. 
The datasets proposed by Xu et al.~(\citeyear{misdataset-xu}) and Zhou et al.~(\citeyear{2023MMD}) only provide descriptions of misconfiguration issues without annotating their root causes. The other dataset~(Wen et al.~\citeyear{2024SlsDetector}) labels the misconfigurations with concrete root cause in code-level. It does not follow a systematic taxonomy of misconfiguration root causes. In our dataset, each case has its own ID, year, software name, link, description, and the type/subtype of root cause.
Finally, since the existing datasets do not provide the reproduced misconfiguration cases, we implemented 6 real-world software misconfigurations and wrapped them in Docker containers so that they can be directly used\footnote{\url{https://github.com/anabioticsoul/misconfiguration\_datasets}}.}

\noindent \textbf{Summary: } 
In total, only {6} out of \TODO{60} studies used publicly available datasets for misconfiguration troubleshooting, among which four studies used the same dataset, two studies used both crafted and real-world misconfiguration errors. 
To complement the existing datasets, in this work, we propose a dataset that contains 772 real-world misconfigurations on seven research targets, along with the root cause and descriptions of these cases.  
Moreover, we also reproduced six misconfiguration cases and wrapped them in Docker containers so that they can be directly used by researchers in this field\footnotemark[\value{footnote}].

\begin{keyfindingbox}{Key Findings of RQ3:}
\begin{itemize}[
  label=$\bullet$,
  leftmargin=*,
  itemsep=0.3em
]

    \item Only \TODO{11} out of \TODO{60} studies have made their misconfiguration troubleshooting tools publicly available. 
    \item The maintenance and adaptability of the tools are problematic. Among the \TODO{11} available tools for misconfiguration troubleshooting, only \TODO{4} tools have received updates in the past three years, and \TODO{9} tools were implemented tightly coupled with their evaluation targets in the papers.
    \item Only {6} out of \TODO{60} studies used publicly available datasets for misconfiguration troubleshooting, among which 4 studies used the same dataset that only contains 98 real-world misconfigurations, one study used the dataset with 84 misconfigurations including crafted ones, and one study used the dataset that consists of 22 multi-option errors.
    \item To complement the existing datasets, we propose a dataset that contains 772 real-world misconfigurations along with the root causes, {subtypes,} and descriptions of these cases. We also reproduced six cases and wrapped them in Docker containers so that they can be directly used.
    
\end{itemize}
\end{keyfindingbox}

\section{RQ4: Challenges and Suggestions}
Synthesizing the findings from both our empirical study (RQ1) and literature analysis (RQ2 \& RQ3), this section identifies the critical gaps remaining in the field. We outline three major challenges and offer actionable suggestions for developers to mitigate these issues.

\subsection{Challenges to troubleshooting misconfigurations}

Combining real-world misconfiguration cases and literature on misconfiguration troubleshooting, in this {subsection,} we summarize the challenges from several perspectives.

\subsubsection{Challenge~1: Insufficient feedback messages for troubleshooting}
One gap between existing troubleshooting tools and real-world misconfiguration cases is due to the quality of error feedback. After carefully reviewing the feedback provided by the questioners in the real-world issues, we found that \textit{the information such as error messages, log outputs, and system prompts is insufficient {for troubleshooting} in {14.51\% (112/772)} of misconfiguration cases in our dataset.} {The error messages provided in the issues were often ambiguous or misleading.} For example, in \textit{RC$\mathit{_{III}}$-Case\#25}, the version of OpenSSL was incompatible with Nginx, but the software only provided a message ``The connection was reset'', which {could not directly help the user solve the problem}. In \textit{RC$\mathit{_{III}}$-Case\#24}, Apache reported a syntax error, but the user finally found that it was caused by an incompatible version of the software. Troubleshooting tools that rely on error feedback usually require detailed debugging information such as error codes, line numbers, or explicit configuration references~(Rabkin and Katz~\citeyear{2011ConfAnalyzer}; Zhang and Ernst~\citeyear{2013confDiagnoser}, \citeyear{2014confSuggester}). They cannot identify the root cause of misconfigurations when only sufficient messages are unavailable.
{In summary, to fix software misconfigurations, one important challenge is to obtain useful information (e.g., error message) for misconfiguration diagnosis. However, many software lack such information, sometimes even provide misleading feedback information unrelated to misconfiguration issues.}

\subsubsection{Challenge~2: Lack of practical misconfiguration troubleshooting tools}
Although much research has been done on misconfiguration troubleshooting, there is still a gap between troubleshooting tools and users when facing real-world misconfiguration issues. In practice, most of the tools are difficult for users to adopt. Specifically, there are three reasons.  
First, most (\TODO{81.67\%}) of the proposed tools are not publicly available. They remain as prototypes developed for academic studies and are not released as open source projects or integrated into production environments. 
Second, most (\TODO{81.82\%}) of the available tools were tightly coupled with the research targets used in the papers, which causes difficulty when applying them to other software. It is still desirable to design a universal or extensible tool that is independent with research targets. 
Finally, most (\TODO{63.64\%}) tools were not continuously maintained and updated. Users can hardly adopt them due to outdated environmental requirements and dependencies. Only \TODO{four} tools for troubleshooting misconfiguration have received updates in the last three years. In summary, these three reasons also make it difficult to conduct a fair evaluation of these tools on the unified misconfiguration benchmark. It underscores the necessity to drive the development of more adaptable troubleshooting tools in the future.

\subsubsection{Challenge~3: Requirement of manual efforts}
Despite progress on misconfiguration diagnosis methods, most existing methods still require manual efforts to diagnose configuration errors. Manual efforts often require that users have a deep understanding of the underlying software, which is unrealistic for most practitioners. Specifically, \textbf{\ding{172}} {Pattern-based methods require manually summarized configuration constraints, {such as the constraints between different configurations}. 
Developers or researchers need to manually analyze source code and documentation to identify the relevant configuration options and their relationships within the system~(Liao et al.~\citeyear{2018manual}; Chen et al.~\citeyear{2020cDep}; Zhang et al.~\citeyear{zhang2021configX}; Zhou et al.~\citeyear{2023MMD}). Manual modeling serves as the foundation for detecting and diagnosing configuration errors, which is labor-intensive and error-prone. In addition, the patterns created by manual efforts often capture very common misconfigurations. It is difficult to extract all potential constraints within the software based on manual modeling. \textbf{\ding{173}} Misconfiguration troubleshooting approaches based on static program analysis also require manual processing. For example, ConfDiagnoser~(Zhang and Ernst~\citeyear{2013confDiagnoser}) and ConfSuggester~(Zhang and Ernst~\citeyear{2014confSuggester}) require users to specify the configuration entry points and associate configuration options with predicates in the source code. Although the technique automates deviation analysis using a profile database, the seed selection for slicing still depends on user input or pre-processing. 
\textbf{\ding{174}} Replay-based methods require manual efforts when reproducing the failure point. Users have to manually trigger the errors while the software system performs dynamic binary instrumentation to trace causal dependencies between the configuration tokens and observed failures or performance anomalies~(Attariyan and Flinn ~\citeyear{2010confAid}, \citeyear{2012Xray}). Similarly, ConfSuggester~(Zhang and Ernst~\citeyear{2014confSuggester}) requires users to provide two versions of the software (before and after behavioral change), manually demonstrate divergent behavior under instrumentation, and mark the configuration files to be monitored. In summary, although existing techniques automate the process of troubleshooting misconfiguration to some extent, many processes still require human intervention. These manual steps, such as specifying configuration constraints, identifying entry points, or reproducing failure conditions, not only require expert knowledge but also limit the scalability and availability of the tools in real-world scenarios.

\subsection{Suggestions}
To facilitate troubleshooting of real-world misconfigurations, we also provide some suggestions for software developers based on the in-depth analysis of real-world issues and the literature in this field.

\subsubsection{Suggestion~1: Improving the transparency of configuration effects to reduce semantic misinterpretation.}

According to our dataset, 33.55\% (259/772) of the misconfigurations involve cases where software functionality deviates from user expectations. Furthermore, 43.24\% (112/259) of these misconfigurations result in silent deviations from expected functionalities without obvious symptoms like crashes. In our collected issues, especially those caused by ambiguous option names (\Cref{rc4.1}), such as \textit{RC$\mathit{_{IV}}$-Case\#4} and \textit{RC$\mathit{_{IV}}$-Case\#120}, users struggled to identify which configuration option should be adjusted to achieve the expected functionality. Therefore, it is essential for software systems to provide log messages that explicitly indicate which configuration options are responsible for the observed behavior. Diagnosis tools that automatically link software error messages to the relevant configuration sections of the documentation would also help address such issues.

\subsubsection{Suggestion~2: Strengthening software checks for external resources and components.}
Environmental configurations (e.g., configurations for software/hardware resources and third-party components) are often overlooked during deployment. However, they account for {48.19\% (372/772)} of all misconfigurations in our dataset. For instance, users may unintentionally specify unavailable resources such as file paths, CPU cores, or network ports, which lead to runtime failures. In other cases, users may configure missing or incompatible external components (e.g., \textit{RC$\mathit{_{III}}$-Case\#24}), causing failures when the system attempts to invoke them. To mitigate such issues, software systems should provide support for checking environmental dependencies during deployment. Specifically, software should verify the availability and compatibility of required resources, external components, and dependencies in advance, and provide explicit feedback when such requirements are not satisfied. Furthermore, we find that {56.9\% (33/58)} of the component incompatibilities were induced by software upgrades. It is suggested that software systems provide tools for configuration migration across versions and recommend the appropriate configuration values for users.

\subsubsection{Suggestion~3: Checking configuration dependencies during user configuring.}
A configuration option may have control or data dependencies with other configuration options. Specifically, configuration dependencies can occur within a single software component (e.g., \textit{RC$\mathit{_{I}}$-Case\#77}) or across components (e.g., \textit{RC$\mathit{_{I}}$-Case\#149}). Violating these dependencies may cause multi-option errors, especially when users modify one configuration option without synchronously updating the correlated options. It is suggested that software systems automatically validate whether related configuration options remain consistent when users configure the software. When dependency violations are detected, the system should provide explicit feedback to indicate which options are inconsistent and how they can be corrected.

\begin{keyfindingbox}{Key Findings of RQ4:}
\begin{itemize}[
  label=$\bullet$,
  leftmargin=*,
  itemsep=0.3em
]

    \item In our dataset, {14.51\%~(112/772)} of the misconfigurations that raised unexpected effects do not have clear error messages for misconfiguration troubleshooting, which makes misconfiguration troubleshooting challenging.
    \item In previous studies, \TODO{81.67\%~(49/60)} of the proposed tools and \TODO{90\%~(54/60)} of the datasets used in the papers are not publicly available, which makes it difficult for researchers to reproduce the experiments and verify the effectiveness of existing methods.
    \item Existing techniques for misconfiguration troubleshooting still rely on manual efforts. For example, program analysis, replay-based and pattern-based approaches for misconfiguration troubleshooting require manual modeling, which brings some obstacles for users. Automation tools for misconfiguration troubleshooting are still desirable.
    
\end{itemize}
\end{keyfindingbox}

\section{Threats to Validity}
The threats to the validity of this work include three aspects.

\noindent\textbf{Real-world misconfiguration collection.} Our dataset was constructed from real-world misconfiguration cases collected from online community Q\&A forums and issue trackers. However, there are still limitations: We did not perform weighted extraction from a small number of software systems, where the number of collected cases varies across systems. The software we analyzed only contains open source software, as they are mature and widely deployed, but may not represent all software types (e.g., commercial software). To mitigate the threat, the collection sources of our dataset covers diverse categories (i.e., development software, web application, database, and cloud software), which ensure the generality of our analysis. In addition, the selected software systems are among the mainstream platforms in common practice, which provides strong representativeness.

\noindent\textbf{Literature collection.} For the systematic literature analysis, we constructed search queries in multiple digital libraries (e.g., IEEE Xplore, ACM DL, ScienceDirect) using the set of keywords, and we applied inclusion and exclusion criteria. However, several limitations remain: Some relevant studies may not have been captured due to publication venue or terminology differences. In addition, we did not include non-English work. To mitigate the threat, our search covered a wide range of high-quality publication venues, ensuring that the majority of influential studies were included. Moreover, we also employed the ``snowballing'' procedure, in which we examined the references and citations of the selected papers to identify and include additional relevant work. Moreover, since English serves as the international academic language, most of the top-tier conferences, journals, and influential research outputs in software engineering are published in English. These limitations may reduce the completeness of our literature analysis, but they do not threaten the validity of our conclusions. 

\noindent\textbf{Data classification and analysis.} The primary threat lies in the manual analysis of the real-world misconfiguration cases. Because categorizing root causes, identifying users' carelessness, and evaluating the insufficiency of feedback messages inherently rely on human comprehension, the results might be subject to researcher bias. To mitigate this threat and ensure the reliability of our findings in RQ1, RQ2 and RQ4, we employed open coding~(Creswell and Poth~\citeyear{creswell2016}) during the data classification phase. Every single issue, including its extracted contextual attributes and assigned root cause, was independently analyzed by two evaluators. Any discrepancies were carefully discussed and resolved with the involvement of a third evaluator. Our systematic analysis effectively minimizes individual bias.

\section{Related Work}
This section places our study within the broader context of software configuration research. We compare our proposed taxonomy with existing empirical studies. We also survey relevant fields such as configuration bug detection and dependency analysis to provide a comprehensive background. Furthermore, we distinguish the scope of this paper from studies on software variability.

\begin{table}[htb]
\footnotesize
\centering
\caption{Comparison of classification with the previous work}
\label{comparison-yin}
\resizebox{\hsize}{!}{
\begin{tabular}{ccccccc|cc}
\toprule
\multirow{3}{*}{Paper} & \multicolumn{6}{c|}{Constraint violation} & \multicolumn{2}{c}{Component integration error} \\
\cline{2-7} \cline{7-9}
& Syntax & Invalid & Misplaced & Duplicate & Intra-component & Cross-component & Component & Component \\ 
& error & option name & configuration & option & error & error & incompatibility & missing \\
\midrule
Yin~et al.~(\citeyear{yin2011empirical}) & \CIRCLE & \CIRCLE & \CIRCLE & \LEFTcircle & \CIRCLE & \Circle & \CIRCLE & \CIRCLE \\
This work & \CIRCLE & \CIRCLE & \CIRCLE & \CIRCLE & \CIRCLE & \CIRCLE & \CIRCLE & \CIRCLE \\
\hline
\multirow{3}{*}{Paper} & \multicolumn{6}{c|}{Resource unavailability} & \multicolumn{2}{c}{Configuration semantic misinterpretation} \\
\cline{2-5} \cline{6-9}
& Resource identifier & Resource & Unauthorized & Hardware & - & - & Ambiguous option & Option value and functional \\
& mismatch & competition & resource access & limitation & - & - & name & requirement mismatch \\
\midrule
Yin~et al.~(\citeyear{yin2011empirical}) & \CIRCLE & \Circle & \Circle & \CIRCLE & - & - & \Circle & \CIRCLE \\
This work & \CIRCLE & \CIRCLE & \CIRCLE & \CIRCLE & - & - & \CIRCLE & \CIRCLE \\
\bottomrule
\end{tabular}
}
\begin{tablenotes}
\scriptsize
\item[] \CIRCLE~ means that the classification fully includes this type of root cause, \LEFTcircle~ means partial inclusion, and \Circle~ means that this type of root cause is not included in the classification.
\end{tablenotes}
\end{table}

\subsection{Empirical studies and literature reviews on software configuration}
Several previous studies have explored misconfigurations through empirical studies and literature reviews. Specifically, Yin~et al.~(\citeyear{yin2011empirical}) analyzed 546 real-world configuration errors, among which 309 cases were from a commercial system. They statistically characterized the observed misconfigurations and summarized the misconfiguration types. The detailed comparison of the classifications proposed by this work and Yin~et al. can be seen in \Cref{comparison-yin}. Compared to their classification, we found four important root causes of software misconfigurations that have not been clearly defined before, i.e., resource competition, unauthorized resource access, cross-component error and ambiguous option name. In addition, we investigate the evolution of research trends in misconfiguration troubleshooting, assess the practicality of existing tools and the availability of misconfiguration datasets, and identify the open challenges that remain unsolved in troubleshooting real-world misconfigurations. Sayagh et al.~(\citeyear{sayagh2018software}) and Le et al.~(\citeyear{le2021surveyonTackling}) conducted literature reviews of software configuration diagnosis and summarized the methods used in the previous works. Sayagh et al. also studied software configuration engineering by interviewing software experts and surveying Java software engineers and pointed out challenges faced by developers. Compared with these works, this paper combines the literature reviews and the causes of real-world cases, aiming at identifying the gap between academic research and user practice in this field. 
There are also other studies trying to narrow the gap between users and software configurations. From the perspective of developers, Zhang et al.~(\citeyear{zhang2021evolutionary}) studied the evolution of configuration design and implementation in cloud systems by analyzing the evolution of source code to figure out how to improve the configuration design. In this paper, we focus on the gap between misconfiguration diagnosis and real-world cases from the perspective of users. Gazzillo et al.~(\citeyear{SPLASH2022together}) proposed a unified conceptual model to bridge different configuration research areas, categorizing the configuration lifecycle into user specifications, developer specifications, and implementations. While our work complements this extensive body of research by providing a fine-grained, empirical analysis of runtime misconfiguration root causes specifically from the troubleshooting perspective in the real world.

\subsection{Configuration bug detection and dependency analysis}
Some studies focus on detecting defects in program or configuration implementations, even when the supplied option values are valid. For instance, Toman and Grossman~(\citeyear{2016staccato}) proposed Staccato, a bug-finding tool designed to detect program logic flaws and state inconsistencies when software processes dynamic configuration updates. Oh et al.~(\citeyear{2021kismet}) introduced kismet, a bug finder that leverages automated theorem proving to statically analyze the Kconfig language and detect unmet dependency bugs. Similarly, Wang et al.~(\citeyear{2023Parachute}) introduced Parachute, a framework aimed at understanding and automatically detecting ``on-the-fly'' configuration bugs that occur during runtime configuration reloading. Other studies focus on extracting configuration dependencies or constraints from source code and log messages. For example, Chen et al.~(\citeyear{2020cDep}) proposed cDep, a static analysis approach to automatically discover and extract complex software configuration dependencies within cloud and datacenter systems. Simon et al.~(\citeyear{2023cfgNet}) developed CfgNet, a framework dedicated to analyzing and tracking equality-based configuration dependencies across different technology stacks. Furthermore, researchers have explored alternative artifacts for constraint extraction. Zhou et al.~(\citeyear{2021confinlog}) proposed ConfInLog, a tool that leverages natural language patterns to automatically infer configuration constraints directly from software log messages. Note that implementations bug detection and configuration dependency analysis fall outside the scope of our systematic literature review.

\subsection{Feature and software variability}
Configuration-related concepts are referred to by multiple terms in both the literature and practice. The scope of what we call configurations in this paper differs from that of software variability. As illustrated by Gazzillo and
Cohen~(\citeyear{SPLASH2022together}), \textit{variability} in Software Product Line (SPL) is typically resolved during the product derivation phase to generate a family of distinct software products from a common codebase. Moreover, variability bugs arise where the selection of an option is valid but triggers the  implementation bug of program. In contrast, our study investigates misconfigurations, where the software implementation is correct, but the user provides an incorrect configuration value, leading to unexpected software behaviors. Some related studies focus on analyzing highly configurable systems and variability-aware implementations. For instance, Liebig et al.~(\citeyear{2013Scalable}) studied scalable analysis strategies for variable software, while von Rhein et al.~(\citeyear{2018Variability}) empirically evaluated variability-aware static analysis techniques at scale. Acher et al.~(\citeyear{2019learning-linux}) built more than 95,000 Linux kernel configurations and found that many build failures were caused by missing tools or dependencies required by specific configuration options rather than actual source-code defects. Other studies focus on variability evolution and debugging. Michelon et al.~(\citeyear{SPLC2021managing}) discussed challenges in maintaining and evolving systems across variants in space and time, and Ngo et al.~(\citeyear{2021Benchmark}) provided a benchmark for variability fault localization in highly configurable systems. In addition, our taxonomy of component integration errors shares similarities with existing work on feature interactions~(Thüm et al.~\citeyear{CSUR2014Classification}; Calder et al.~\citeyear{CN2003feature-interaction}), as both address unintended behaviors arising from the complex dependencies between different parts of a system. However, a component is different from a feature. Features are user-visible units of product functionality. By contrast, the errors we study arise from dependencies among software components, external dependencies, or third-party libraries, which are not product features. For example, a version mismatch between a database driver and the database server, or a missing dependency library, is a component integration problem rather than a feature interaction. While these studies provide important insights into feature interaction, variability management, and implementation-level variability bugs, they differ from our work.

\section{Conclusion}
In this study, we conducted an empirical study on real-world configuration errors and literature analysis on software misconfigurations.
First, we constructed a real-world misconfiguration dataset, consisting of 772 misconfiguration issues collected from online platforms. By manually analyzing each case and performing cross-validation, We summarized the root causes of misconfigurations into four types, i.e., constraint violation, resource unavailability, component integration error, and configuration semantic misinterpretation. We clearly defined each type of root cause and classified them into several subtypes to assist misconfiguration diagnosis.
Second, we systematically reviewed the research papers on misconfiguration troubleshooting published during the past two decades to investigate whether existing tools can effectively address real-world configuration issues. We found that despite research progress, users still face many challenges in diagnosing software misconfigurations, especially the lack of practical tools that can be adopted directly. We analyzed the trends of existing literature on misconfiguration troubleshooting, summarized the challenges faced by users, and highlighted the suggestions for mitigating and diagnosing misconfigurations. We released the dataset for follow-up research. We believe that this paper can benefit researchers, developers, and users in the field of software misconfigurations.

\subsection*{Acknowledgements}
We sincerely thank the reviewers for their constructive comments and valuable suggestions, which helped improve the quality and clarity of this paper. We also thank Zongshang Shen from Nankai University for his assistance in constructing part of the reproduced cases in this study.

\subsection*{Author contributions}
Yuhao Liu: Conceptualization, methodology, data collection, data analysis, visualization, writing the original draft, and editing the manuscript. Yingnan Zhou: Data collection, data analysis, and editing the manuscript. Hanfeng Zhang: Data collection, data analysis, and writing the original draft. Zhiwei Chang: Data collection, data analysis, and writing the original draft. Sihan Xu: Project administration, conceptualization, methodology, validation, editing the manuscript, and funding acquisition. Yan Jia: Project administration, conceptualization, methodology, validation, data collection, data analysis, editing the manuscript, and funding acquisition. Wei Wang: Supervision, reviewing the manuscript, and funding acquisition. Juncheng Hu: Supervision, reviewing the manuscript. Zheli Liu: Project administration, supervision, reviewing the manuscript, and funding acquisition.

\subsection*{Funding}
This work was supported by the National Natural Science Foundation of China (under Grants 62572258, 62441227, U22B2027), the Joint Fund of the National Natural Science Foundation of China (under Grant U25B2028), the Key Program of the National Natural Science Foundation of China (under Grant 62432012), the National Natural Science Foundation of Tianjin, China (under Grant 25JCQNJC01380), and the Fundamental Research Funds for the Central Universities (under Grants 079-63263257 and 079-63261161).

\subsection*{Data availability}
The dataset of 772 real-world misconfiguration cases used in this study is publicly available\footnote{\url{https://github.com/anabioticsoul/misconfiguration\_datasets}}.

\section*{Declarations}

\subsection*{Ethical approval}
This article does not contain any studies with human participants or animals performed by any of the authors.

\subsection*{Informed consent}
Not applicable.

\subsection*{Conflicts of interest}
The authors declare that they have no conflict of interest.

\subsection*{Clinical trial number}
Not applicable.

\bibliographystyle{spbasic}      
\bibliography{reference}

@misc{misdataset-xu,
  author = "Tianyin Xu",
  year  = 2015,
  title = "Misconfiguration Dataset",
  note = "\url{https://github.com/tianyin/configuration_datasets/tree/master/configissues}"
}

@misc{misdataset-wen,
  author = "Jinfeng Wen",
  year  = 2024,
  title = "Evaluation Dataset",
  note = "\url{https://github.com/WenJinfeng/SlsDetector_ConfigurationDetection/tree/main/EvaluationDataset}"
}

@article{CN2003feature-interaction,
  title={Feature interaction: a critical review and considered forecast},
  author={Calder, Muffy and Kolberg, Mario and Magill, Evan H and Reiff-Marganiec, Stephan},
  journal={Computer Networks},
  volume={41},
  number={1},
  pages={115--141},
  year={2003},
  publisher={Elsevier}
}

@article{CSUR2014Classification,
  title={A Classification and Survey of Analysis Strategies for Software Product Lines},
  author={Thomas Th{\"u}m and Sven Apel and Christian K{\"a}stner and Ina Schaefer and Gunter Saake},
  journal={ACM Computing Surveys (CSUR)},
  year={2014},
  volume={47},
  pages={1 - 45},
  url={https://api.semanticscholar.org/CorpusID:9766972}
}

@inproceedings{SPLC2021managing,
  author       = {Gabriela Karoline Michelon and
                  David Obermann and
                  Wesley K. G. Assun{\c{c}}{\~{a}}o and
                  Lukas Linsbauer and
                  Paul Gr{\"{u}}nbacher and
                  Alexander Egyed},
  editor       = {Mohammad Reza Mousavi and
                  Pierre{-}Yves Schobbens},
  title        = {Managing systems evolving in space and time: four challenges for maintenance,
                  evolution and composition of variants},
  booktitle    = {{SPLC} '21: 25th {ACM} International Systems and Software Product
                  Line Conference, Leicester, United Kingdom, September 6-11, 2021,
                  Volume {A}},
  pages        = {75--80},
  publisher    = {{ACM}},
  year         = {2021},
  url          = {https://doi.org/10.1145/3461001.3461660},
  doi          = {10.1145/3461001.3461660},
  bibsource    = {dblp computer science bibliography, https://dblp.org}
}

@inproceedings{2021Benchmark,
  title={Variability fault localization: A benchmark},
  author={Ngo, Kien-Tuan and Nguyen, Thu-Trang and Nguyen, Son and Vo, Hieu Dinh},
  booktitle={Proceedings of the 25th ACM International Systems and Software Product Line Conference-Volume A},
  pages={120--125},
  year={2021}
}

@article{SPLASH2022together,
  title={Bringing Together Configuration Research: Towards a Common Ground},
  author={Paul Gazzillo and Myra B. Cohen},
  journal={Proceedings of the 2022 ACM SIGPLAN International Symposium on New Ideas, New Paradigms, and Reflections on Programming and Software},
  year={2022},
  url={https://api.semanticscholar.org/CorpusID:253305336}
}

@phdthesis{2019learning-linux,
  title={Learning from thousands of build failures of Linux kernel configurations},
  author={Acher, Mathieu and Martin, Hugo and Pereira, Juliana Alves and Blouin, Arnaud and Khelladi, Djamel Eddine and J{\'e}z{\'e}quel, Jean-Marc},
  year={2019},
  school={Inria; IRISA}
}

@inproceedings{2021kismet,
  author       = {Jeho Oh and
                  Necip Fazil Yildiran and
                  Julian Braha and
                  Paul Gazzillo},
  editor       = {Diomidis Spinellis and
                  Georgios Gousios and
                  Marsha Chechik and
                  Massimiliano Di Penta},
  title        = {Finding broken Linux configuration specifications by statically analyzing
                  the Kconfig language},
  booktitle    = {{ESEC/FSE} '21: 29th {ACM} Joint European Software Engineering Conference
                  and Symposium on the Foundations of Software Engineering, Athens,
                  Greece, August 23-28, 2021},
  pages        = {893--905},
  publisher    = {{ACM}},
  year         = {2021},
  url          = {https://doi.org/10.1145/3468264.3468578},
  doi          = {10.1145/3468264.3468578},
  bibsource    = {dblp computer science bibliography, https://dblp.org}
}

@inproceedings{2013Scalable,
  author       = {J{\"{o}}rg Liebig and
                  Alexander von Rhein and
                  Christian K{\"{a}}stner and
                  Sven Apel and
                  Jens D{\"{o}}rre and
                  Christian Lengauer},
  editor       = {Bertrand Meyer and
                  Luciano Baresi and
                  Mira Mezini},
  title        = {Scalable analysis of variable software},
  booktitle    = {Joint Meeting of the European Software Engineering Conference and
                  the {ACM} {SIGSOFT} Symposium on the Foundations of Software Engineering,
                  ESEC/FSE'13, Saint Petersburg, Russian Federation, August 18-26, 2013},
  pages        = {81--91},
  publisher    = {{ACM}},
  year         = {2013},
  url          = {https://doi.org/10.1145/2491411.2491437},
  doi          = {10.1145/2491411.2491437},
  bibsource    = {dblp computer science bibliography, https://dblp.org}
}

@article{2018Variability,
  title={Variability-Aware Static Analysis at Scale},
  author={Alexander von Rhein and J{\"o}rg Liebig and Andreas Janker and Christian K{\"a}stner and Sven Apel},
  journal={ACM Transactions on Software Engineering and Methodology (TOSEM)},
  year={2018},
  volume={27},
  pages={1 - 33},
  url={https://api.semanticscholar.org/CorpusID:52840340}
}

@article{2023cfgNet,
  author       = {Sebastian Simon and
                  Nicolai Ruckel and
                  Norbert Siegmund},
  title        = {CfgNet: {A} Framework for Tracking Equality-Based Configuration Dependencies
                  Across a Software Project},
  journal      = {{IEEE} Trans. Software Eng.},
  volume       = {49},
  number       = {8},
  pages        = {3955--3971},
  year         = {2023},
  url          = {https://doi.org/10.1109/TSE.2023.3274349},
  doi          = {10.1109/TSE.2023.3274349},
  bibsource    = {dblp computer science bibliography, https://dblp.org}
}

@inproceedings{2008CURE,
  author       = {Jules White and
                  Douglas C. Schmidt and
                  David Benavides and
                  Pablo Trinidad and
                  Antonio Ruiz Cort{\'{e}}s},
  title        = {Automated Diagnosis of Product-Line Configuration Errors in Feature Models},
  booktitle    = {Software Product Lines, 12th International Conference, {SPLC} 2008,
                  Limerick, Ireland, September 8-12, 2008, Proceedings},
  pages        = {225--234},
  publisher    = {{IEEE} Computer Society},
  year         = {2008},
  url          = {https://doi.org/10.1109/SPLC.2008.16},
  doi          = {10.1109/SPLC.2008.16},
  bibsource    = {dblp computer science bibliography, https://dblp.org}
}

@article{2010CURE-JSS,
  author       = {Jules White and
                  David Benavides and
                  Douglas C. Schmidt and
                  Pablo Trinidad and
                  Brian Dougherty and
                  Antonio Ruiz Cort{\'{e}}s},
  title        = {Automated diagnosis of feature model configurations},
  journal      = {J. Syst. Softw.},
  volume       = {83},
  number       = {7},
  pages        = {1094--1107},
  year         = {2010},
  url          = {https://doi.org/10.1016/j.jss.2010.02.017},
  doi          = {10.1016/J.JSS.2010.02.017},
  bibsource    = {dblp computer science bibliography, https://dblp.org}
}

@inproceedings{2013SmartFixer,
  author       = {Bo Wang and
                  Leonardo Teixeira Passos and
                  Yingfei Xiong and
                  Krzysztof Czarnecki and
                  Haiyan Zhao and
                  Wei Zhang},
  editor       = {Tomoji Kishi and
                  Stan Jarzabek and
                  Stefania Gnesi},
  title        = {SmartFixer: fixing software configurations based on dynamic priorities},
  booktitle    = {17th International Software Product Line Conference, {SPLC} 2013,
                  Tokyo, Japan - August 26 - 30, 2013},
  pages        = {82--90},
  publisher    = {{ACM}},
  year         = {2013},
  url          = {https://doi.org/10.1145/2491627.2491640},
  doi          = {10.1145/2491627.2491640},
  bibsource    = {dblp computer science bibliography, https://dblp.org}
}

@inproceedings{yin2011empirical,
  title={An empirical study on configuration errors in commercial and open source systems},
  author={Yin, Zuoning and Ma, Xiao and Zheng, Jing and Zhou, Yuanyuan and Bairavasundaram, Lakshmi N and Pasupathy, Shankar},
  booktitle={Proceedings of the Twenty-Third ACM Symposium on Operating Systems Principles},
  pages={159--172},
  year={2011},
  note = {\url{https://doi.org/10.1145/2043556.2043572}}
}

@inproceedings{2010confAid,
  title={Automating Configuration Troubleshooting with Dynamic Information Flow Analysis.},
  author={Attariyan, Mona and Flinn, Jason},
  booktitle={OSDI},
  volume={10},
  number={2010},
  pages={1--14},
  year={2010}
}

@inproceedings{2010sherlog,
  title={Sherlog: error diagnosis by connecting clues from run-time logs},
  author={Yuan, Ding and Mai, Haohui and Xiong, Weiwei and Tan, Lin and Zhou, Yuanyuan and Pasupathy, Shankar},
  booktitle={Proceedings of the fifteenth International Conference on Architectural support for programming languages and operating systems},
  pages={143--154},
  year={2010},
  note = {\url{https://doi.org/10.1145/1736020.1736038}}
}

@inproceedings{2011ConfAnalyzer,
  title={Precomputing possible configuration error diagnoses},
  author={Rabkin, Ariel and Katz, Randy},
  booktitle={2011 26th IEEE/ACM International Conference on Automated Software Engineering (ASE 2011)},
  pages={193--202},
  year={2011},
  note = {\url{https://doi.org/10.1109/ASE.2011.6100053}}
}

@inproceedings{2011ConfAnalyzer_issta,
  title={Using static analysis to diagnose configuration errors},
  author={Rabkin, Ariel and Katz, Randy},
  booktitle={Proc. of the 11th Int’l Symp. on Software Testing and Analysis (ISSTA)},
  year={2011}
}

@inproceedings{2012Xray,
  title={X-ray: Automating root-cause diagnosis of performance anomalies in production software},
  author={Attariyan, Mona and Chow, Michael and Flinn, Jason},
  booktitle={10th $\{$USENIX$\}$ Symposium on Operating Systems Design and Implementation ($\{$OSDI$\}$ 12)},
  pages={307--320},
  year={2012}
}

@inproceedings{2013SPEX,
  title={Do not blame users for misconfigurations},
  author={Xu, Tianyin and Zhang, Jiaqi and Huang, Peng and Zheng, Jing and Sheng, Tianwei and Yuan, Ding and Zhou, Yuanyuan and Pasupathy, Shankar},
  booktitle={Proceedings of the Twenty-Fourth ACM Symposium on Operating Systems Principles},
  pages={244--259},
  year={2013},
  note = {\url{https://doi.org/10.1145/2517349.2522727}}
}

@inproceedings{2013confDiagnoser,
  title={Automated diagnosis of software configuration errors},
  author={Zhang, Sai and Ernst, Michael D},
  booktitle={2013 35th International Conference on Software Engineering (ICSE)},
  pages={312--321},
  year={2013},
  note = {\url{https://doi.org/10.1109/ICSE.2013.6606577}}
}

@inproceedings{2013confdiagnoser_3page,
  title={Confdiagnoser: An automated configuration error diagnosis tool for java software},
  author={Zhang, Sai},
  booktitle={2013 35th International Conference on Software Engineering (ICSE)},
  pages={1438--1440},
  year={2013},
  note = {\url{https://doi.org/10.1109/ICSE.2013.6606737}}
}

@inproceedings{2013confDebugger,
  title={Automated diagnosis of software misconfigurations based on static analysis},
  author={Dong, Zhen and Ghanavati, Mohammadreza and Andrzejak, Artur},
  booktitle={2013 IEEE International Symposium on Software Reliability Engineering Workshops (ISSREW)},
  pages={162--168},
  year={2013},
  note = {\url{https://doi.org/10.1109/ISSREW.2013.6688897}},
}

@inproceedings{2014confSuggester,
  title={Which configuration option should i change?},
  author={Zhang, Sai and Ernst, Michael D},
  booktitle={Proceedings of the 36th International Conference on Software Engineering},
  pages={152--163},
  year={2014},
  note = {\url{https://doi.org/10.1145/2568225.2568251}}
}

@inproceedings{2015confDoctor,
  title={Practical and accurate pinpointing of configuration errors using static analysis},
  author={Dong, Zhen and Andrzejak, Artur and Shao, Kun},
  booktitle={2015 IEEE International Conference on Software Maintenance and Evolution (ICSME)},
  pages={171--180},
  year={2015},
  note = {\url{https://doi.org/10.1109/ICSM.2015.7332463}}
}

@inproceedings{2016PCHECK,
  title={Early detection of configuration errors to reduce failure damage},
  author={Xu, Tianyin and Jin, Xinxin and Huang, Peng and Zhou, Yuanyuan and Lu, Shan and Jin, Long and Pasupathy, Shankar},
  booktitle={12th $\{$USENIX$\}$ Symposium on Operating Systems Design and Implementation ($\{$OSDI$\}$ 16)},
  pages={619--634},
  year={2016}
}

@inproceedings{2017Cross-stack,
  title={On cross-stack configuration errors},
  author={Sayagh, Mohammed and Kerzazi, Noureddine and Adams, Bram},
  booktitle={2017 IEEE/ACM 39th International Conference on Software Engineering (ICSE)},
  pages={255--265},
  year={2017},
  note = {\url{https://doi.org/10.1109/ICSE.2017.31}}
}

@inproceedings{2020cDep,
  title={Understanding and discovering software configuration dependencies in cloud and datacenter systems},
  author={Chen, Qingrong and Wang, Teng and Legunsen, Owolabi and Li, Shanshan and Xu, Tianyin},
  booktitle={Proceedings of the 28th ACM Joint Meeting on European Software Engineering Conference and Symposium on the Foundations of Software Engineering},
  pages={362--374},
  note = {\url{https://doi.org/10.1145/3368089.3409727}},
  year={2020}
}

@article{zhang2021configX,
  title={Static detection of silent misconfigurations with deep interaction analysis},
  author={Zhang, Jialu and Piskac, Ruzica and Zhai, Ennan and Xu, Tianyin},
  journal={Proceedings of the ACM on Programming Languages},
  volume={5},
  number={OOPSLA},
  pages={1--30},
  year={2021},
  publisher={ACM New York, NY, USA},
  note = {\url{https://doi.org/10.1145/3485517}}
}

@article{ding2023real,
  title={Real-Time Diagnosis of Configuration Errors for Software of AI Server Infrastructure},
  author={Xu, Guangquan and Ding, Xinru and Xu, Sihan and Jia, Yan and Liu, Shaoying and Feng, Shicheng and Zheng, Xi},
  journal={IEEE Transactions on Dependable and Secure Computing},
  year={2023},
  publisher={IEEE},
  note = {\url{https://doi.org/10.1109/TDSC.2023.3266007}}
}

@inproceedings{2023Parachute,
  title={Understanding and detecting on-the-fly configuration bugs},
  author={Wang, Teng and Jia, Zhouyang and Li, Shanshan and Zheng, Si and Yu, Yue and Xu, Erci and Peng, Shaoliang and Liao, Xiangke},
  booktitle={2023 IEEE/ACM 45th International Conference on Software Engineering (ICSE)},
  pages={628--639},
  year={2023},
  organization={IEEE}
}

@article{2023MMD,
  title={Multi-misconfiguration Diagnosis via Identifying Correlated Configuration Parameters},
  author={Zhou, Yingnan and Hu, Xue and Xu, Sihan and Jia, Yan and Liu, Yuhao and Wang, Junyong and Xu, Guangquan and Wang, Wei and Liu, Shaoying and Baker, Thar},
  journal={IEEE Transactions on Software Engineering},
  year={2023},
  publisher={IEEE},
  note = {\url{https://doi.org/10.1109/TSE.2023.3308755}}
}

@inproceedings{2023DiagConfig,
  author       = {Zhiming Chen and
                  Pengfei Chen and
                  Peipei Wang and
                  Guangba Yu and
                  Zilong He and
                  Genting Mai},
  editor       = {Satish Chandra and
                  Kelly Blincoe and
                  Paolo Tonella},
  title        = {DiagConfig: Configuration Diagnosis of Performance Violations in Configurable
                  Software Systems},
  booktitle    = {Proceedings of the 31st {ACM} Joint European Software Engineering
                  Conference and Symposium on the Foundations of Software Engineering,
                  {ESEC/FSE} 2023, San Francisco, CA, USA, December 3-9, 2023},
  pages        = {566--578},
  publisher    = {{ACM}},
  year         = {2023},
  url          = {https://doi.org/10.1145/3611643.3616300},
  doi          = {10.1145/3611643.3616300},
  bibsource    = {dblp computer science bibliography, https://dblp.org}
}

@inproceedings{2024LogConfigLocalizer,
  author       = {Shiwen Shan and
                  Yintong Huo and
                  Yuxin Su and
                  Yichen Li and
                  Dan Li and
                  Zibin Zheng},
  editor       = {Maria Christakis and
                  Michael Pradel},
  title        = {Face It Yourselves: An LLM-Based Two-Stage Strategy to Localize Configuration
                  Errors via Logs},
  booktitle    = {Proceedings of the 33rd {ACM} {SIGSOFT} International Symposium on
                  Software Testing and Analysis, {ISSTA} 2024, Vienna, Austria, September
                  16-20, 2024},
  pages        = {13--25},
  publisher    = {{ACM}},
  year         = {2024},
  url          = {https://doi.org/10.1145/3650212.3652106},
  doi          = {10.1145/3650212.3652106},
  bibsource    = {dblp computer science bibliography, https://dblp.org}
}

@inproceedings{2020Violet,
  author       = {Yigong Hu and
                  Gongqi Huang and
                  Peng Huang},
  title        = {Automated Reasoning and Detection of Specious Configuration in Large
                  Systems with Symbolic Execution},
  booktitle    = {14th {USENIX} Symposium on Operating Systems Design and Implementation,
                  {OSDI} 2020, Virtual Event, November 4-6, 2020},
  pages        = {719--734},
  publisher    = {{USENIX} Association},
  year         = {2020},
  url          = {https://www.usenix.org/conference/osdi20/presentation/hu},
  bibsource    = {dblp computer science bibliography, https://dblp.org}
}

@inproceedings{2016staccato,
  title={Staccato: A bug finder for dynamic configuration updates},
  author={Toman, John and Grossman, Dan},
  booktitle={30th European Conference on Object-Oriented Programming (ECOOP 2016)},
  year={2016},
  organization={Schloss Dagstuhl-Leibniz-Zentrum fuer Informatik}
}

@article{2018manual,
  title={Do you really know how to configure your software? configuration constraints in source code may help},
  author={Liao, Xiangke and Zhou, Shulin and Li, Shanshan and Jia, Zhouyang and Liu, Xiaodong and He, Haochen},
  journal={IEEE Transactions on Reliability},
  volume={67},
  number={3},
  pages={832--846},
  year={2018},
  publisher={IEEE},
  note = {\url{https://doi.org/10.1109/TR.2018.2834419}}
}

@inproceedings{2021confinlog,
  title={ConfInLog: Leveraging Software Logs to Infer Configuration Constraints},
  author={Zhou, Shulin and Liu, Xiaodong and Li, Shanshan and Jia, Zhouyang and Zhang, Yuanliang and Wang, Teng and Li, Wang and Liao, Xiangke},
  booktitle={2021 IEEE/ACM 29th International Conference on Program Comprehension (ICPC)},
  pages={94--105},
  year={2021},
  organization={IEEE}
}

@article{2003peerpressure,
  title={Peerpressure: A statistical method for automatic misconfiguration troubleshooting},
  author={Wang, Helen J and Platt, John and Chen, Yu and Zhang, Ruyun and Wang, Yi-Min},
  journal={Technical report, Microsoft Research},
  year={2003},
  publisher={Citeseer}
}

@inproceedings{2004peerpressure-osdi,
  author       = {Helen J. Wang and
                  John C. Platt and
                  Yu Chen and
                  Ruyun Zhang and
                  Yi{-}Min Wang},
  editor       = {Eric A. Brewer and
                  Peter Chen},
  title        = {Automatic Misconfiguration Troubleshooting with PeerPressure},
  booktitle    = {6th Symposium on Operating System Design and Implementation {(OSDI}
                  2004), San Francisco, California, USA, December 6-8, 2004},
  pages        = {245--258},
  publisher    = {{USENIX} Association},
  year         = {2004},
  url          = {http://www.usenix.org/events/osdi04/tech/wang.html},
  bibsource    = {dblp computer science bibliography, https://dblp.org}
}

@article{2004strider,
  title={Strider: A black-box, state-based approach to change and configuration management and support},
  author={Wang, Yi-Min and Verbowski, Chad and Dunagan, John and Chen, Yu and Wang, Helen J and Yuan, Chun and Zhang, Zheng},
  journal={Science of Computer Programming},
  volume={53},
  number={2},
  pages={143--164},
  year={2004},
  publisher={Elsevier},
  note = {\url{https://doi.org/10.1016/j.scico.2003.12.009}}
}

@inproceedings{2004Chronus,
  title={Configuration Debugging as Search: Finding the Needle in the Haystack.},
  author={Whitaker, Andrew and Cox, Richard S and Gribble, Steven D and others},
  booktitle={OSDI},
  volume={4},
  pages={6--6},
  year={2004}
}

@inproceedings{2004state-based,
  title={Combining High Level Symptom Descriptions and Low Level State Information for Configuration Fault Diagnosis.},
  author={Lao, Ni and Wen, Ji-Rong and Ma, Wei-Ying and Wang, Yi-Min},
  booktitle={LISA},
  volume={4},
  pages={151--158},
  year={2004}
}

@article{2006EventTraceBased,
  title={Automated known problem diagnosis with event traces},
  author={Yuan, Chun and Lao, Ni and Wen, Ji-Rong and Li, Jiwei and Zhang, Zheng and Wang, Yi-Min and Ma, Wei-Ying},
  journal={ACM SIGOPS Operating Systems Review},
  volume={40},
  number={4},
  pages={375--388},
  year={2006},
  publisher={ACM New York, NY, USA},
  note = {\url{https://doi.org/10.1145/1217935.1217972}}
}

@inproceedings{2007snitch,
  title={Snitch: Interactive decision trees for troubleshooting misconfigurations},
  author={Mickens, James and Szummer, Martin and Narayanan, Dushyanth},
  booktitle={In Proceedings of the 2007 Workshop on Tackling Computer Systems Problems with Machine Learning Techniques},
  year={2007}
}

@article{2007autobash,
  author = {Ya{-}Yunn Su and
                  Mona Attariyan and
                  Jason Flinn},
  editor = {Thomas C. Bressoud and
                  M. Frans Kaashoek},
  title = {AutoBash: improving configuration management with operating system
                  causality analysis},
  booktitle = {Proceedings of the 21st {ACM} Symposium on Operating Systems Principles
                  2007, {SOSP} 2007, Stevenson, Washington, USA, October 14-17, 2007},
  journal = {ACM SIGOPS Operating Systems Review},
  volume = {41},
  number={6},
  pages={237--250},
  year={2007},
  publisher={ACM New York, NY, USA},
  note = {\url{https://doi.org/10.1145/1294261.1294284}}
}

@inproceedings{2008signature-based,
  title={Automatic Software Fault Diagnosis by Exploiting Application Signatures.},
  author={Ding, Xiaoning and Huang, Hai and Ruan, Yaoping and Shaikh, Anees and Zhang, Xiaodong},
  booktitle={LISA},
  volume={8},
  pages={23--39},
  year={2008}
}

@inproceedings{2008sigConf,
  title={Using Causality to Diagnose Configuration Bugs.},
  author={Attariyan, Mona and Flinn, Jason},
  booktitle={USENIX Annual Technical Conference},
  pages={281--286},
  year={2008}
}

@inproceedings{2009predicate-based,
  author       = {Ya{-}Yunn Su and
                  Jason Flinn},
  editor       = {Geoffrey M. Voelker and
                  Alec Wolman},
  title        = {Automatically Generating Predicates and Solutions for Configuration
                  Troubleshooting},
  booktitle    = {Proceedings of the 2009 {USENIX} Annual Technical Conference, {USENIX}
                  {ATC} 2009, San Diego, CA, USA, June 14-19, 2009},
  publisher    = {{USENIX} Association},
  year         = {2009},
}

@inproceedings{2011CODE,
  title={Context-based online configuration-error detection},
  author={Yuan, Ding and Xie, Yinglian and Panigrahy, Rina and Yang, Junfeng and Verbowski, Chad and Kumar, Arunvijay},
  booktitle={Proceedings of the 2011 USENIX conference on USENIX annual technical conference},
  pages={28--28},
  year={2011}
}

@inproceedings{2011machineLear-based,
  title={Capturing expert knowledge for automated configuration fault diagnosis},
  author={Wang, Mengliao and Shi, Xiaoyu and Wong, Kenny},
  booktitle={2011 IEEE 19th International Conference on Program Comprehension},
  pages={205--208},
  year={2011},
  note = {\url{https://doi.org/10.1109/ICPC.2011.24}}
}

@inproceedings{2012Cloud-basedMisDetec,
  title={Misconfiguration detection for cloud datacenters using decision tree analysis},
  author={Uchiumi, Tetsuya and Kikuchi, Shinji and Matsumoto, Yasuhide},
  booktitle={2012 14th Asia-Pacific Network Operations and Management Symposium (APNOMS)},
  pages={1--4},
  year={2012},
  note = {\url{https://doi.org/10.1109/APNOMS.2012.6356072}}
}

@inproceedings{2013Cloud-based,
  title={A differential approach for configuration fault localization in cloud environments},
  author={Kannan, Kalapriya and Bhamidipaty, Anuradha},
  booktitle={2013 IEEE International Conference on Cloud Engineering (IC2E)},
  pages={250--257},
  year={2013},
  note = {\url{https://doi.org/10.1109/IC2E.2013.51}}
}

@inproceedings{2014Encore,
  title={Encore: Exploiting system environment and correlation information for misconfiguration detection},
  author={Zhang, Jiaqi and Renganarayana, Lakshminarayanan and Zhang, Xiaolan and Ge, Niyu and Bala, Vasanth and Xu, Tianyin and Zhou, Yuanyuan},
  booktitle={Proceedings of the 19th international conference on Architectural support for programming languages and operating systems},
  pages={687--700},
  year={2014},
  note = {\url{https://doi.org/10.1145/2541940.2541983}}
}

@inproceedings{2015confvalley,
  title={Confvalley: A systematic configuration validation framework for cloud services},
  author={Huang, Peng and Bolosky, William J and Singh, Abhishek and Zhou, Yuanyuan},
  booktitle={Proceedings of the Tenth European Conference on Computer Systems},
  pages={1--16},
  year={2015},
  note = {\url{https://doi.org/10.1145/2741948.2741963}}
}

@inproceedings{2016configc,
  title={Probabilistic automated language learning for configuration files},
  author={Santolucito, Mark and Zhai, Ennan and Piskac, Ruzica},
  booktitle={International Conference on Computer Aided Verification},
  pages={80--87},
  year={2016},
  note = {\url{https://doi.org/10.1007/978-3-319-41540-6\_5}}
}

@inproceedings{2017dexter,
  title={Dexter: faster troubleshooting of misconfiguration cases using system logs},
  author={Talwadker, Rukma},
  booktitle={Proceedings of the 10th ACM International Systems and Storage Conference},
  pages={1--12},
  year={2017},
  note = {\url{https://doi.org/10.1145/3078468.3078484}}
}

@article{2017configv,
  title={Synthesizing configuration file specifications with association rule learning},
  author={Santolucito, Mark and Zhai, Ennan and Dhodapkar, Rahul and Shim, Aaron and Piskac, Ruzica},
  journal={Proceedings of the ACM on Programming Languages},
  volume={1},
  number={OOPSLA},
  pages={1--20},
  year={2017},
  publisher={ACM New York, NY, USA},
  note = {\url{https://doi.org/10.1145/3133888}}
}

@inproceedings{2020rex,
  title={Rex: Preventing bugs and misconfiguration in large services using correlated change analysis},
  author={Mehta, Sonu and Bhagwan, Ranjita and Kumar, Rahul and Bansal, Chetan and Maddila, Chandra and Ashok, B and Asthana, Sumit and Bird, Christian and Kumar, Aditya},
  booktitle={17th $\{$USENIX$\}$ Symposium on Networked Systems Design and Implementation ($\{$NSDI$\}$ 20)},
  pages={435--448},
  year={2020}
}

@inproceedings{2020pracextractor,
  title={Pracextractor: Extracting configuration good practices from manuals to detect server misconfigurations},
  author={Xiang, Chengcheng and Huang, Haochen and Yoo, Andrew and Zhou, Yuanyuan and Pasupathy, Shankar},
  booktitle={2020 $\{$USENIX$\}$ Annual Technical Conference ($\{$USENIX$\}$$\{$ATC$\}$ 20)},
  pages={265--280},
  year={2020}
}

@article{2020machineLear-based3,
  title={Predicting Hadoop misconfigurations using machine learning},
  author={Robert, Andrew and Gupta, Apaar and Shenoy, Vinayak and Sitaram, Dinkar and Kalambur, Subramaniam},
  journal={Software: Practice and Experience},
  volume={50},
  number={7},
  pages={1168--1183},
  year={2020},
  publisher={Wiley Online Library},
  note = {\url{https://doi.org/10.1002/spe.2790}}
}

@inproceedings{2021ConfDetect,
  title={Software misconfiguration troubleshooting based on state analysis},
  author={Li, Ke and Xue, Yuan and Shao, Yujie and Su, Bing and Tan, Yu-an and Hu, Jingjing},
  booktitle={2021 IEEE Sixth International Conference on Data Science in Cyberspace (DSC)},
  pages={361--366},
  year={2021},
  note = {\url{https://doi.org/10.1109/DSC53577.2021.00057}}
}

@article{2024SlsDetector,
  title={LLM-Based Misconfiguration Detection for AWS Serverless Computing},
  author={Wen, Jinfeng and Chen, Zhenpeng and Sarro, Federica and Zhu, Zixi and Liu, Yi and Ping, Haodi and Wang, Shangguang},
  journal={arXiv preprint arXiv:2411.00642},
  year={2024},
  note = {\url{https://doi.org/10.48550/arXiv.2411.00642}}
}

@inproceedings{xu2015knobs,
  title={Hey, you have given me too many knobs!: Understanding and dealing with over-designed configuration in system software},
  author={Xu, Tianyin and Jin, Long and Fan, Xuepeng and Zhou, Yuanyuan and Pasupathy, Shankar and Talwadker, Rukma},
  booktitle={Proceedings of the 2015 10th Joint Meeting on Foundations of Software Engineering},
  pages={307--319},
  year={2015},
  note = {\url{https://doi.org/10.1145/2786805.2786852}}
}

@article{2014stackoverflow,
  title={What are developers talking about? an analysis of topics and trends in stack overflow},
  author={Barua, Anton and Thomas, Stephen W and Hassan, Ahmed E},
  journal={Empirical software engineering},
  volume={19},
  pages={619--654},
  year={2014},
  note = {\url{https://doi.org/10.1007/s10664-012-9231-y}},
  publisher={Springer}
}

@inproceedings{wang2024exploratory,
  title={An Exploratory Investigation of Log Anomalies in Unmanned Aerial Vehicles},
  author={Wang, Dinghua and Li, Shuqing and Xiao, Guanping and Liu, Yepang and Sui, Yulei and He, Pinjia and Lyu, Michael R},
  booktitle={Proceedings of the IEEE/ACM 46th International Conference on Software Engineering},
  pages={1--13},
  year={2024},
  note = {\url{https://doi.org/10.1145/3597503.3639186}}
}

@article{1977measurement,
  title={The measurement of observer agreement for categorical data},
  author={Landis, J Richard and Koch, Gary G},
  journal={biometrics},
  pages={159--174},
  year={1977},
  publisher={JSTOR},
  note = {\url{https://doi.org/10.2307/2529310}}
}

@techreport{2007SLRguidelines,
  title={Guidelines for performing systematic literature reviews in software engineering},
  author={Keele, Staffs and others},
  year={2007},
  institution={Technical report, ver. 2.3 ebse technical report. ebse}
}

@article{2015SMSguidelines,
  author       = {Kai Petersen and
                  Sairam Vakkalanka and
                  Ludwik Kuzniarz},
  title        = {Guidelines for conducting systematic mapping studies in software engineering:
                  An update},
  journal      = {Inf. Softw. Technol.},
  volume       = {64},
  pages        = {1--18},
  year         = {2015},
  url          = {https://doi.org/10.1016/j.infsof.2015.03.007},
  doi          = {10.1016/J.INFSOF.2015.03.007},
  bibsource    = {dblp computer science bibliography, https://dblp.org}
}

@inproceedings{2014snowballing,
  title={Guidelines for snowballing in systematic literature studies and a replication in software engineering},
  author={Wohlin, Claes},
  booktitle={Proceedings of the 18th international conference on evaluation and assessment in software engineering},
  pages={1--10},
  year={2014},
  note = {\url{https://doi.org/10.1145/2601248.2601268}}
}

@book{creswell2016,
  title={Qualitative inquiry and research design: Choosing among five approaches},
  author={Creswell, John W and Poth, Cheryl N},
  year={2016},
  publisher={Sage publications}
}

@article{Empirical2015Xu,
  title={Systems approaches to tackling configuration errors: A survey},
  author={Xu, Tianyin and Zhou, Yuanyuan},
  journal={ACM Computing Surveys (CSUR)},
  volume={47},
  number={4},
  pages={1--41},
  year={2015},
  publisher={ACM New York, NY, USA},
  note = {\url{https://doi.org/10.1145/2791577}}
}

@inproceedings{wang2021exploratory,
  title={An exploratory study of autopilot software bugs in unmanned aerial vehicles},
  author={Wang, Dinghua and Li, Shuqing and Xiao, Guanping and Liu, Yepang and Sui, Yulei},
  booktitle={Proceedings of the 29th ACM Joint Meeting on European Software Engineering Conference and Symposium on the Foundations of Software Engineering},
  pages={20--31},
  year={2021},
  note = {\url{https://doi.org/10.1145/3468264.3468559}}
}

@inproceedings{zhang2021evolutionary,
  title={An Evolutionary Study of Configuration Design and Implementation in Cloud Systems},
  author={Zhang, Yuanliang and He, Haochen and Legunsen, Owolabi and Li, Shanshan and Dong, Wei and Xu, Tianyin},
  booktitle={2021 IEEE/ACM 43rd International Conference on Software Engineering (ICSE)},
  pages={188--200},
  year={2021},
  note = {\url{https://doi.org/10.1109/ICSE43902.2021.00029}}
}

@article{sayagh2018software,
  title={Software configuration engineering in practice interviews, survey, and systematic literature review},
  author={Sayagh, Mohammed and Kerzazi, Noureddine and Adams, Bram and Petrillo, Fabio},
  journal={IEEE Transactions on Software Engineering},
  volume={46},
  number={6},
  pages={646--673},
  year={2018},
  publisher={IEEE},
  note = {\url{https://doi.org/10.1109/TSE.2018.2867847}}
}

@inproceedings{2013confeagle,
  title={Confeagle: Automated analysis of configuration vulnerabilities in web applications},
  author={Eshete, Birhanu and Villafiorita, Adolfo and Weldemariam, Komminist and Zulkernine, Mohammad},
  booktitle={2013 IEEE 7th International Conference on Software Security and Reliability},
  pages={188--197},
  year={2013},
  note = {\url{https://doi.org/10.1109/SERE.2013.30}}
}

@inproceedings{2021white-box,
  author       = {Max Weber and
                  Sven Apel and
                  Norbert Siegmund},
  title        = {White-Box Performance-Influence Models: {A} Profiling and Learning
                  Approach},
  booktitle    = {43rd {IEEE/ACM} International Conference on Software Engineering,
                  {ICSE} 2021, Madrid, Spain, 22-30 May 2021},
  pages        = {1059--1071},
  publisher    = {{IEEE}},
  year         = {2021},
  url          = {https://doi.org/10.1109/ICSE43902.2021.00099},
  doi          = {10.1109/ICSE43902.2021.00099},
  bibsource    = {dblp computer science bibliography, https://dblp.org}
}

@inproceedings{han2021confprof,
  title={ConfProf: White-Box Performance Profiling of Configuration Options},
  author={Han, Xue and Yu, Tingting and Pradel, Michael},
  booktitle={Proceedings of the ACM/SPEC International Conference on Performance Engineering},
  pages={1--8},
  year={2021},
  note = {\url{https://doi.org/10.1145/3427921.3450255}}
}

@inproceedings{li2020learnconf,
  title={Statically inferring performance properties of software configurations},
  author={Li, Chi and Wang, Shu and Hoffmann, Henry and Lu, Shan},
  booktitle={Proceedings of the Fifteenth European Conference on Computer Systems},
  pages={1--16},
  year={2020},
  note = {\url{https://doi.org/10.1145/3342195.3387520}}
}

@inproceedings{mahgoub2019sophia,
  title={$\{$SOPHIA$\}$: Online reconfiguration of clustered nosql databases for time-varying workloads},
  author={Mahgoub, Ashraf and Wood, Paul and Medoff, Alexander and Mitra, Subrata and Meyer, Folker and Chaterji, Somali and Bagchi, Saurabh},
  booktitle={2019 $\{$USENIX$\}$ Annual Technical Conference ($\{$USENIX$\}$$\{$ATC$\}$ 19)},
  pages={223--240},
  year={2019}
}

@article{2020configcrusher,
  title={Configcrusher: Towards white-box performance analysis for configurable systems},
  author={Velez, Miguel and Jamshidi, Pooyan and Sattler, Florian and Siegmund, Norbert and Apel, Sven and K{\"a}stner, Christian},
  journal={Automated Software Engineering},
  volume={27},
  pages={265--300},
  year={2020},
  publisher={Springer},
  note = {\url{https://doi.org/10.1007/s10515-020-00273-8}}
}

@article{hsu2018scout,
  title={Scout: An experienced guide to find the best cloud configuration},
  author={Hsu, Chin-Jung and Nair, Vivek and Menzies, Tim and Freeh, Vincent W},
  journal={arXiv preprint arXiv:1803.01296},
  year={2018},
  note = {\url{https://doi.org/10.48550/arXiv.1803.01296}}
}

@article{wang2018smartconf,
  title={Understanding and auto-adjusting performance-sensitive configurations},
  author={Wang, Shu and Li, Chi and Hoffmann, Henry and Lu, Shan and Sentosa, William and Kistijantoro, Achmad Imam},
  journal={ACM SIGPLAN Notices},
  volume={53},
  number={2},
  pages={154--168},
  year={2018},
  publisher={ACM New York, NY, USA},
  note = {\url{https://doi.org/10.1145/3173162.3173206}}
}

@inproceedings{wang2022agilectrl,
  title={AgileCtrl: a self-adaptive framework for configuration tuning},
  author={Wang, Shu and Hoffmann, Henry and Lu, Shan},
  booktitle={Proceedings of the 30th ACM Joint European Software Engineering Conference and Symposium on the Foundations of Software Engineering},
  pages={459--471},
  year={2022},
  note = {\url{https://doi.org/10.1145/3540250.3549136}}
}

@inproceedings{he2022safetune,
  title={Multi-intention-aware configuration selection for performance tuning},
  author={He, Haochen and Jia, Zhouyang and Li, Shanshan and Yu, Yue and Zhou, Chenglong and Liao, Qing and Wang, Ji and Liao, Xiangke},
  booktitle={Proceedings of the 44th International Conference on Software Engineering},
  pages={1431--1442},
  year={2022},
  note = {\url{https://doi.org/10.1145/3510003.3510094}}
}

@inproceedings{2008configre,
  title={Towards automatic reverse engineering of software security configurations},
  author={Wang, Rui and Wang, XiaoFeng and Zhang, Kehuan and Li, Zhuowei},
  booktitle={Proceedings of the 15th ACM conference on Computer and communications security},
  pages={245--256},
  year={2008},
  note = {\url{https://doi.org/10.1145/1455770.1455802}}
}

@inproceedings{le2021surveyonTackling,
  title={A Survey on Tackling Software Configuration Faults},
  author={Le, Wenge and Wang, Yong and Yang, Fei and Wang, Xue and Wang, Shouhang},
  booktitle={2021 7th International Symposium on System and Software Reliability (ISSSR)},
  pages={11--17},
  year={2021},
  note = {\url{https://doi.org/10.1109/ISSSR53171.2021.00036}}
}

@article{2021ConfigMiner,
  author       = {Mohammed Sayagh and
                  Ahmed E. Hassan},
  title        = {ConfigMiner: Identifying the Appropriate Configuration Options for
                  Config-Related User Questions by Mining Online Forums},
  journal      = {{IEEE} Trans. Software Eng.},
  volume       = {47},
  number       = {12},
  pages        = {2907--2918},
  year         = {2021},
  url          = {https://doi.org/10.1109/TSE.2020.2973997},
  doi          = {10.1109/TSE.2020.2973997},
  bibsource    = {dblp computer science bibliography, https://dblp.org}
}

@incollection{soot,
  title={Soot: A Java bytecode optimization framework},
  author={Vall{\'e}e-Rai, Raja and Co, Phong and Gagnon, Etienne and Hendren, Laurie and Lam, Patrick and Sundaresan, Vijay},
  booktitle={CASCON First Decade High Impact Papers},
  pages={214--224},
  year={2010},
  publisher    = {{IBM}},
  note = {\url{https://doi.org/10.1145/367845.368008}}
}

@article{tucek2007triage,
  title={Triage: diagnosing production run failures at the user's site},
  author={Tucek, Joseph and Lu, Shan and Huang, Chengdu and Xanthos, Spiros and Zhou, Yuanyuan},
  journal={ACM SIGOPS Operating Systems Review},
  volume={41},
  number={6},
  pages={131--144},
  year={2007},
  publisher={ACM New York, NY, USA},
  note = {\url{https://doi.org/10.1145/1294261.1294275}}
}





\noindent
\begin{minipage}[t]{0.18\textwidth}
    \vspace{0pt}
    \includegraphics[width=\linewidth]{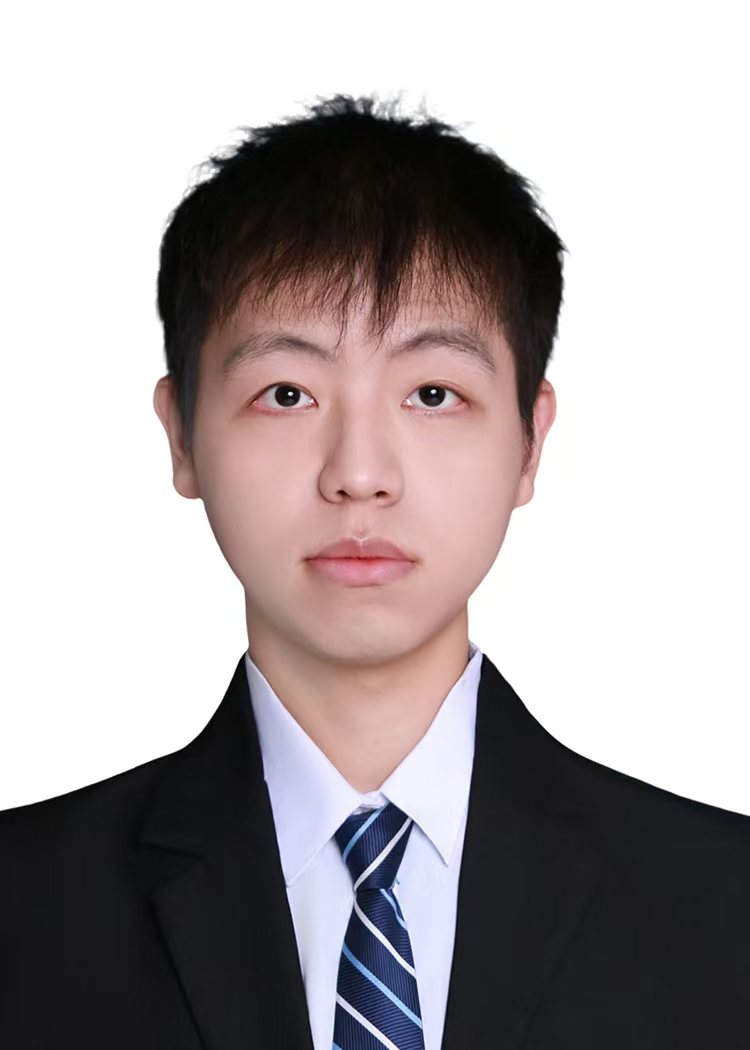}
\end{minipage}
\hfill
\begin{minipage}[t]{0.78\textwidth}
    \vspace{0pt}
    \textbf{Yuhao~Liu} is a Ph.D. student at the College of Cryptography and Cyber Science, Nankai University, Tianjin, China, since 2023. He received his M.S. degree from the School of Computer and Information Technology, Beijing Jiaotong University, China. His research focuses on software security, and his interests include programming languages, reverse engineering, and cloud-native security.
\end{minipage}

\vspace{1em}

\noindent
\begin{minipage}[t]{0.18\textwidth}
    \vspace{0pt}
    \includegraphics[width=\linewidth]{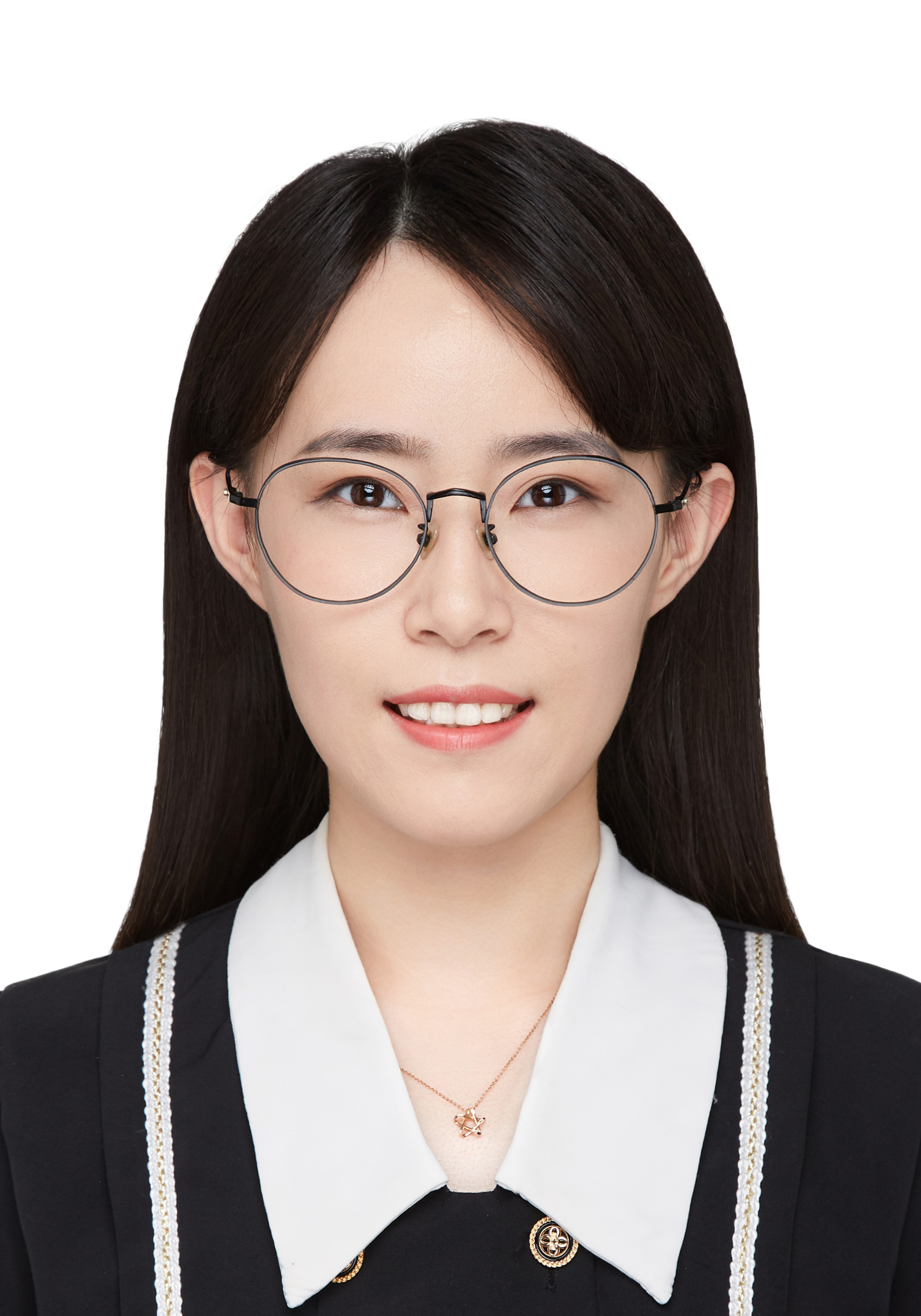}
\end{minipage}
\hfill
\begin{minipage}[t]{0.78\textwidth}
    \vspace{0pt}
    \textbf{Yingnan~Zhou} is a Ph.D. student at Nankai University, Tianjin, China, majoring in cyber security. Her research interests include configuration security, program analysis, and software security.
\end{minipage}

\vspace{1em}

\noindent
\begin{minipage}[t]{0.18\textwidth}
    \vspace{0pt}
    \includegraphics[width=\linewidth]{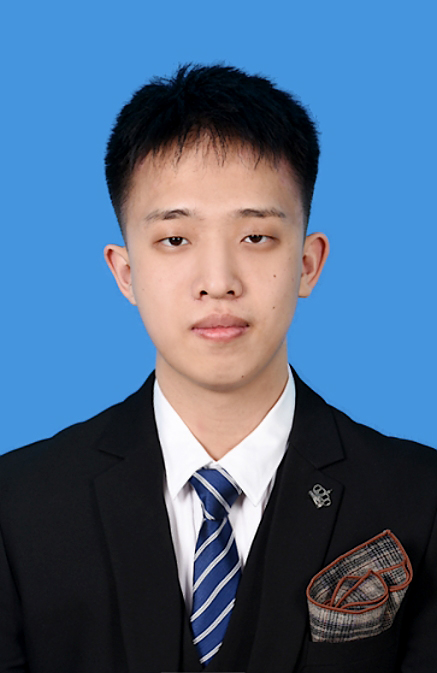}
\end{minipage}
\hfill
\begin{minipage}[t]{0.78\textwidth}
    \vspace{0pt}
    \textbf{Hanfeng~Zhang} received his M.S. degree from the School of Cyber Science, Nankai University, Tianjin, China, in 2024. He is currently a game client developer at TiMi Studio Group, Tencent. His research interests include software security and fuzzing. He aims to enhance the security of consumer-facing software through continuous research in the fields of software security and software engineering.
\end{minipage}

\vspace{1em}

\noindent
\begin{minipage}[t]{0.18\textwidth}
    \vspace{0pt}
    \includegraphics[width=\linewidth]{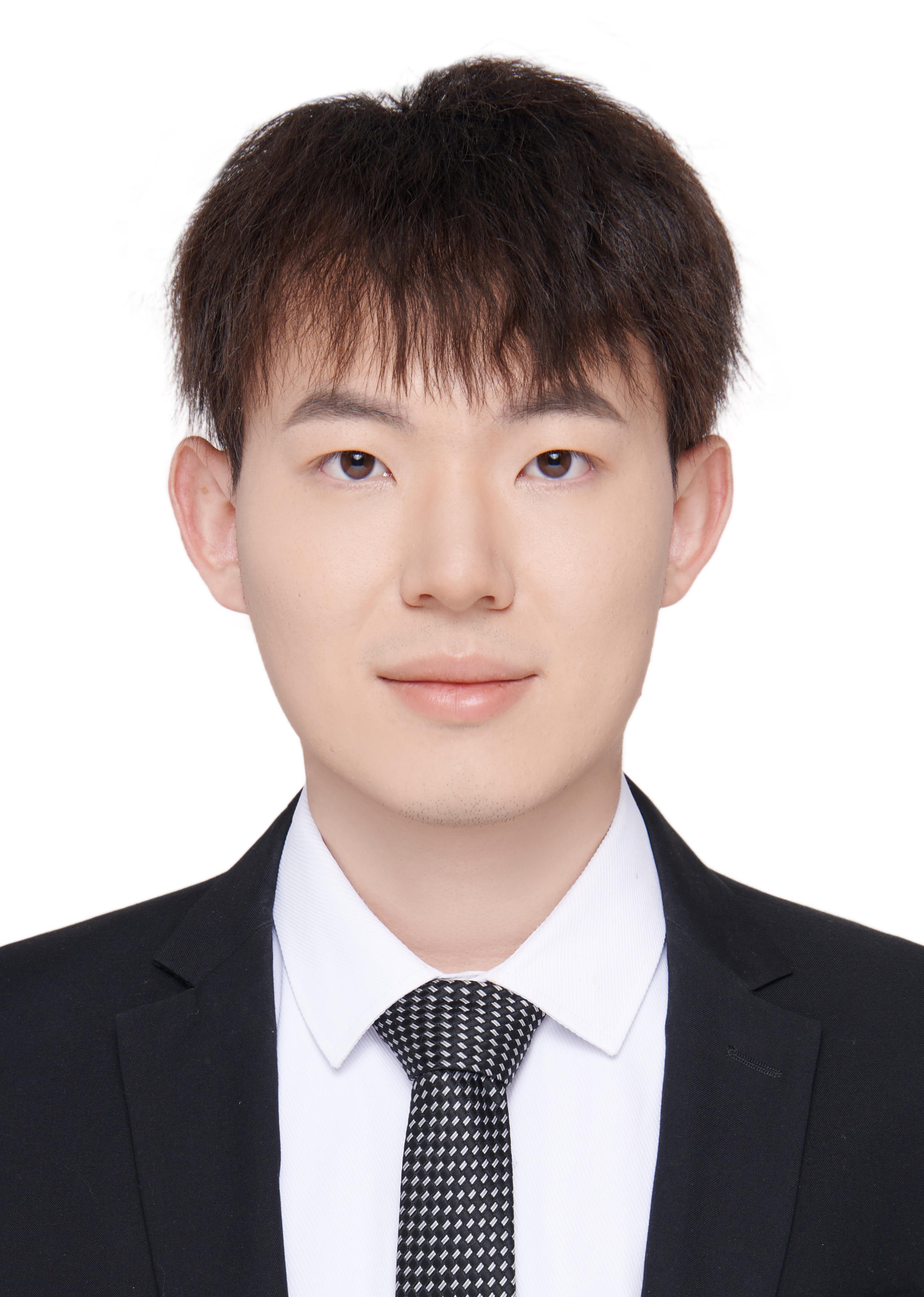}
\end{minipage}
\hfill
\begin{minipage}[t]{0.78\textwidth}
    \vspace{0pt}
    \textbf{Zhiwei~Chang} received the master’s degree from the College of Cyber Science, Nankai University, Tianjin, China, in 2024. His research interests include software security and software configuration vulnerabilities.
\end{minipage}

\vspace{1em}

\noindent
\begin{minipage}[t]{0.18\textwidth}
    \vspace{0pt}
    \includegraphics[width=\linewidth]{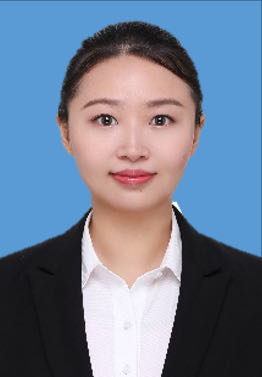}
\end{minipage}
\hfill
\begin{minipage}[t]{0.78\textwidth}
    \vspace{0pt}
    \textbf{Sihan~Xu} received the B.Sc. and Ph.D. degrees in computer science from Nankai University in 2013 and 2018, respectively. For her research, she spent a year with the National University of Singapore. She is currently an associate professor with College of Cryptography and Cyber Science, Nankai University. Her research interests include intelligent software engineering and AI security.
\end{minipage}

\vspace{1em}

\noindent
\begin{minipage}[t]{0.18\textwidth}
    \vspace{0pt}
    \includegraphics[width=\linewidth]{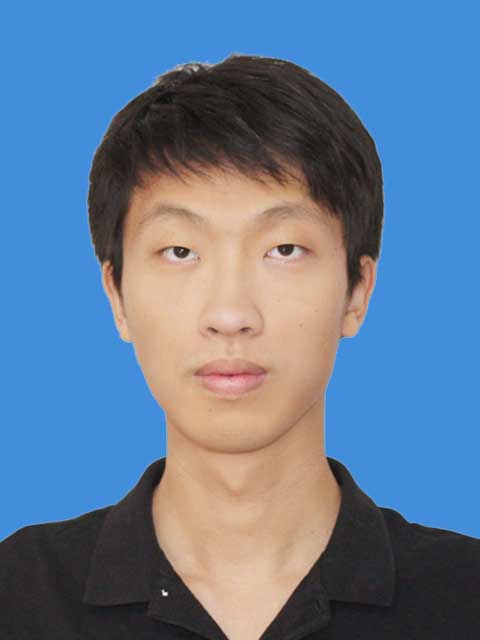}
\end{minipage}
\hfill
\begin{minipage}[t]{0.78\textwidth}
    \vspace{0pt}
    \textbf{Yan~Jia} received the Ph.D. degree from the School of Cyber Engineering, Xidian University, Xi’an, China, in 2020. He is a Research Associate with the College of Cryptography and Cyber Science, Nankai University, Tianjin, China. His interests include discovering and understanding new design or logic security vulnerabilities in real-world systems, including IoT, Web/browser, mobile, and network. His work helped many high-profile vendors improve their products’ security, including Amazon, Microsoft, Apple, and Google.
\end{minipage}

\vspace{1em}

\noindent
\begin{minipage}[t]{0.18\textwidth}
    \vspace{0pt}
    \includegraphics[width=\linewidth]{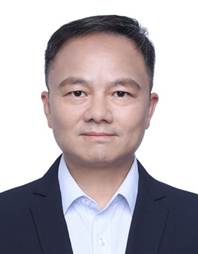}
\end{minipage}
\hfill
\begin{minipage}[t]{0.78\textwidth}
    \vspace{0pt}
    \textbf{Wei~Wang} is a full Professor with Ministry of Education Key Lab for Intelligent Networks and Network Security, aka MOE KLINNS Lab, Xi’an Jiaotong University, China. He is also an adjunct Professor with school of computer science and technology, Beijing Jiaotong University. He received the Ph.D. degree from Xi'an Jiaotong University, in 2006. He was a Post-Doctoral Researcher with the University of Trento, Italy, from 2005 to 2006. He was a Post-Doctoral Researcher with TELECOM Bretagne and with INRIA, France, from 2007 to 2008. He was also a European ERCIM Fellow with the Norwegian University of Science and Technology (NTNU), Norway, and with the Interdisciplinary Centre for Security, Reliability, and Trust (SnT), University of Luxembourg, from 2009 to 2011. He was a faculty member with Beijing Jiaotong University from 2011 to 2024. His recent research interests lie in privacy-preserving computation and blockchain He has authored or co-authored over 100 peer-reviewed articles in various journals and international conferences, including IEEE TDSC, IEEE TIFS, IEEE TSE, USENIX Security, ACM CCS, AAAI, Ubicomp, IEEE INFOCOM. He received the ACM CCS 2023 Distinguished Paper Award. He is an Elsevier ``highly cited Chinese Researchers''. He is an Associate Editor for IEEE Transactions on Dependable and Secure Computing, and an Editorial Board Member of Computers \& Security and of Frontiers of Computer Science. He is a vice chair of ACM SIGSAC China.
\end{minipage}

\vspace{1em}

\noindent
\begin{minipage}[t]{0.18\textwidth}
    \vspace{0pt}
    \includegraphics[width=\linewidth]{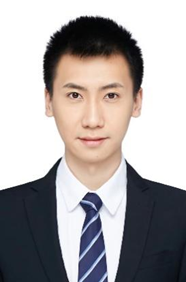}
\end{minipage}
\hfill
\begin{minipage}[t]{0.78\textwidth}
    \vspace{0pt}
    \textbf{Juncheng~Hu} received his bachelor’s degree and doctor of engineering degree from Jilin University in 2017 and 2022. He is currently a associate professor at Jilin University. His research interests include computer architecture, storage engine, trusted computing for big data, and AI security.
\end{minipage}

\vspace{1em}

\noindent
\begin{minipage}[t]{0.18\textwidth}
    \vspace{0pt}
    \includegraphics[width=\linewidth]{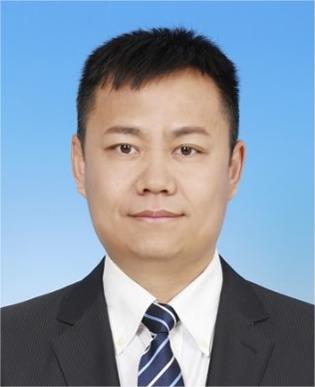}
\end{minipage}
\hfill
\begin{minipage}[t]{0.78\textwidth}
    \vspace{0pt}
    \textbf{Zheli~Liu}
received the BSc and MSc degrees in computer science and a PhD degree in computer application from Jilin University, China, in 2002, 2005, and 2009, respectively. After a postdoctoral fellowship with Nankai University, he joined the College of Cyber Science, Nankai University, in 2011. He is currently a professor at Nankai University. His research interests include applied cryptography and data privacy protection.
\end{minipage}



\end{document}